\documentclass[preprint,12pt]{elsarticle}

\usepackage{amssymb}

\usepackage{amsmath}
\usepackage{appendix}

\newcommand{\beq}{\begin{equation}}
\newcommand{\eeq}{\end{equation}}
\newcommand{\bea}{\begin{align}}
\newcommand{\eea}{\end{align}}

\newcommand{\topt}{T_O}

\newcommand{\e}{\eta}
\newcommand{\om}{\omega}
\newcommand{\omi}{\omega}
\newcommand{\w}{\Omega}
\newcommand{\wi}{\Omega}
\newcommand{\x}{\xi}
\newcommand{\xz}{\xi^0}
\newcommand{\xb}{\bar{\xi}}

\newcommand{\var}{\mathrm{Var}}
\newcommand{\cov}{\mathrm{Cov}}
\newcommand{\fln}{\lfloor n/2 \rfloor}
\newcommand{\sign}{\mathrm{sign}}
\newcommand{\smo}{\Sigma_\mu}
\newcommand{\smb}{\Sigma_{\bar{\mu}}}
\newcommand{\smq}{\Sigma_q}

\journal{...}

\begin{document}

\begin{frontmatter}



\title{Properties of neutrality tests based on allele frequency spectrum}


\author[uab,crag]{L. Ferretti\corref{cor1}}
\ead{luca.ferretti@uab.cat}
\author[japan]{G. Marmorini}
\author[uab,crag]{S. Ramos-Onsins}                                   
\cortext[cor1]{Corresponding author}
\address[uab]{Department of Animal Science and Food, Facultat de Veterinaria, Universitat Autonoma de Barcelona, 08193 Bellaterra, Spain}
\address[crag]{Centre for Research in Agricultural Genomics (CRAG), 08193 Bellaterra, Spain}
\address[japan]{Department of Physics, Keio University, 223-8521
Kanagawa, Yokohama, Hiyoshi 4-1-1, Japan}

\begin{abstract}
One of the main necessities for population geneticists 
is the availability of statistical tools that
enable to accept or reject the neutral Wright-Fisher model with high power. A number of statistical tests have been developed
to detect specific deviations from the null frequency spectrum in different directions (i.e., Tajima's D, Fu and
Li's F and D test, Fay and Wu's H). Recently, a general framework was proposed to generate all neutrality tests that are linear functions of the frequency spectrum. In this framework, a family of optimal tests was developed to have almost maximum power against a specific alternative evolutionary scenario. Following these developments, in this paper we provide a thorough discussion of linear and nonlinear neutrality tests. First, we present the general framework for linear tests and emphasize the importance of the property of
scalability with the sample size (that is, the results of the tests should not depend on the sample size), which, if missing, can guide to
errors in data interpretation. The motivation and structure of linear optimal tests are discussed. In a further generalization, we develop a general framework for nonlinear neutrality tests and we derive nonlinear optimal tests for polynomials of any degree in the frequency spectrum.


\end{abstract}

\begin{keyword}
Coalescent theory, Site frequency spectrum, Population genetics, Statistical power, Summary statistics

\end{keyword}

\end{frontmatter}


\tableofcontents

\section{Introduction}
\label{intro}

Statistical tests for neutrality are important and useful tools for population genetics. Since the development of molecular genetics
techniques allowed to obtain nucleotide sequences for the study of populations genetics \cite{kreitman1983nucleotide}, a number of
neutrality tests have been developed with the objective to facilitate the interpretation of an increasing volume of molecular data.
Statistical tests for neutrality have been generated exploiting the different properties of the stationary neutral model. Examples of tests
are the HKA \cite{hudson1987test}, which takes advantage of the polymorphism/divergence relationship across independent loci in a multilocus
framework, and the Lewontin-Krakauer test \cite{lewontin1973distribution}, which looks for an unexpected level of population differentiation
in a locus in relation to other loci. Also, there is another family of tests related to linkage disequilibrium, as the one developed by
\cite{sabeti2002detecting}, which detect long haplotypes at unusual elevated frequencies in candidate regions. 

An important family of these tests, often used as summary statistics, is built on the frequency spectrum of allele polymorphisms. This family includes the well known tests by Tajima \cite{tajima1989statistical}, Fu and Li \cite{fu1993statistical} and Fay and Wu \cite{fay2000hitchhiking}. If an outgroup is available, these tests are based on the unfolded spectrum $\xi_i$, that is, the number of segregating sites with a derived allele frequency of $i$ in a sample of (haploid) size $n$. Without an outgroup, it is not possible to distinguish derived and ancestral alleles and the only available data correspond to the folded spectrum $\eta_i$, that is, the number of segregating sites with a minor allele frequency of $i$. The quantities $\xi_i$ and $\eta_i$ are all positive and the range of allele frequencies is $1\leq i\leq n-1$ for the unfolded spectrum, $1\leq i\leq \fln$ for the folded spectrum. The average spectra for the standard Wright-Fisher neutral model are given by
\begin{equation}
E(\xi_i)=\frac{1}{i}\theta L\quad , \quad E(\eta_i)=\frac{n}{i(n-i)(1+\delta_{i,n-i})}\theta L\ ,\label{vev}
\end{equation}
where $L$ is the length of the sequence and $\theta=2p\mu N_e$, where $\mu$ is the mutation rate, $p$ is the ploidy and $N_e$ is the effective population size\footnote{Note that we define $\theta$ as the rescaled mutation rate per base and not per sequence. 
Apart from this,  
we follow the notation of \cite{achaz2009frequency} and \cite{fu1995statistical}.}. Note that the spectra are proportional to $\theta$.

In a recent paper by Achaz \cite{achaz2009frequency}, a general framework for these tests was presented. The general tests proposed there were of the form 
\begin{equation}
T_\w=\frac{\sum_{i=1}^{n-1}i\w_i\x_i}{\sqrt{\var\left( \sum_{j=1}^{n-1}j\w_j\x_j\right)}}\quad ,\quad T^*_\w=\frac{\sum_{i=1}^{\fln}i\w_i^*\e_i}{\sqrt{\var\left( \sum_{j=1}^{\fln}j\w^*_j\e_j\right)}}
\end{equation}
that are centered (i.e., they have a null expectation value) if the weights $\w_i,\w_i^*$ satisfy the conditions $\sum_{i=1}^{n-1}\w_i=0$ and $\sum_{i=1}^{\fln}\w^*_i=0$.  
This framework allows the construction of many new neutrality tests and has been used to obtain optimal tests for specific alternative scenarios \cite{ferretti2010optimal}. However the original framework does not take into account the dependence of the tests on the sample size, as emphasized in \cite{ferretti2010optimal}. It is important to choose this dependence in order to obtain results that are as independent as possible on sample size. Moreover, the framework presented in \cite{achaz2009frequency} covers just a large subfamily of neutrality tests based on the frequency spectrum, that is, the class of tests that are linear functions of the spectrum. This subfamily contains almost all the tests that can be found in the literature with the exception of the $G_\xi,G_\eta$ tests of Fu \cite{fu1996new}, which are second order polynomials in the spectrum whose form is related with Hotelling statistics. Since these $G_\xi,G_\eta$ tests were shown to be quite effective in some scenarios, it would be interesting to build a general framework for these quadratic (and more generally nonlinear) tests.

In this paper we provide a detailed study of the properties of the whole family of tests based on allele frequency spectrum, beginning with the discussion of the most interesting case, i.e., linear tests. We present a thorough analysis of a simple proposal for the scaling of the tests with the sample size, then we analyze the geometrical properties of the optimal tests presented in \cite{ferretti2010optimal} and we propose generalizations of $D'$ test to general linear tests and linear optimal tests. Finally, we go beyond the framework presented in \cite{achaz2009frequency} and discuss the most general class of tests, that is, polynomials of any order in $\xi_i,\eta_i$, and obtain the optimal tests for polynomials of any order. These results allow to build new and more effective tests. The proofs can be found in Appendix \ref{proofs}.

\section{Linear neutrality tests}

\subsection{General framework}

As discussed by Achaz \cite{achaz2009frequency}, the general form for linear tests based on the unfolded spectrum can be written as
\beq
T_\w=
\frac{\sum_{i=1}^{n-1}i\w_i\x_i}{\sqrt{\var\left( \sum_{j=1}^{n-1}j\w_j\x_j\right)}}\label{testgeno}
\eeq
where $\w_i$ is a set of weights satisfying the condition
\beq
\sum_{i=1}^{n-1}\w_i=0\ . \label{sumw}
\eeq
This is the most general form if we require that the test is centered and with variance $1$, that is, $E(T_\w)=0$ and $\var(T_\w)=1$. The condition of centeredness can be obtained substituting the spectrum with its average in the standard neutral model, given by the equations (\ref{vev}).

Alternatively, it is sufficient to choose any pair of unbiased estimators of $\theta$ based on the unfolded spectrum
\beq
\hat{\theta}_\om=\frac{1}{L}\sum_{i=1}^{n-1}i\om_i\x_i\quad ,\quad \hat{\theta}_{\om'}=\frac{1}{L}\sum_{i=1}^{n-1}i\om'_i\x_i \label{esttheta}
\eeq
with weights $\om_i,\om'_i$ that obey the conditions
\beq
\sum_{i=1}^{n-1}\om_i=1\quad ,\quad \sum_{i=1}^{n-1}\om'_i=1
\eeq
to obtain a new test for neutrality: 
\beq
T_\w=\frac{\hat{\theta}_\om-\hat{\theta}_{\om'}}{\sqrt{\var (\hat{\theta}_\om-\hat{\theta}_{\om'})}}=
\frac{\sum_{i=1}^{n-1}i(\om_i-\om'_i)\x_i}{\sqrt{\var\left(
\sum_{j=1}^{n-1}j(\om_j-\om'_j)\x_j\right)}}=\frac{\sum_{i=1}^{n-1}i\w_i\x_i}{\sqrt{\var\left(
\sum_{j=1}^{n-1}j\w_j\x_j\right)}} \label{testgen} 
\eeq
that is equivalent to the definition (\ref{testgeno}) with $\w_i=\om_i-\om'_i$. Therefore a test $T_\w$ is defined by real vectors $\w$ or $\om,\om'$ satisfying the above normalization conditions. 

If an outgroup is not available, then the test should be based on the folded spectrum and has the general form: 
\begin{align}
T^*_\w=&\frac{\hat{\theta}^*_\om-\hat{\theta}^*_{\om'}}{\sqrt{\var (\hat{\theta}^*_\om-\hat{\theta}^*_{\om'})}}=\frac{\sum_{i=1}^{\fln}
i(n-i)(1+\delta_{n,2i})
(\om^*_i-\om^{*\prime}_i)\e_i}{\sqrt{\var\left( \sum_{j=1}^{\fln}
j(n-j)(1+\delta_{n,2j})
(\om^*_j-\om^{*\prime}_j)\e_j\right)}}=\nonumber \\ =&\frac{\sum_{i=1}^{\fln}
i(n-i)(1+\delta_{n,2i})
\w_i^*\e_i}{\sqrt{\var\left( \sum_{j=1}^{\fln}
j(n-j)(1+\delta_{n,2j})
\w^*_j\e_j\right)}}
\end{align}
where the weights $\w^*_i=\om^*_i-\om^{*\prime}_i$ satisfy the conditions
\beq
\sum_{i=1}^{\fln}\om^*_i=1\quad ,\quad \sum_{i=1}^{\fln}\om^{*\prime}_i=1\qquad \Rightarrow 
\qquad \sum_{i=1}^{\fln}\w^*_i=0
\eeq
and
\beq
\hat{\theta}^*_\om=\frac{1}{L}\sum_{i=1}^{\fln}\frac{i(n-i)(1+\delta_{n,2i})}{n}\om^*_i\e_i\quad ,\quad \hat{\theta}^*_{\om'}=\frac{1}{L}\sum_{i=1}^{\fln}\frac{i(n-i)(1+\delta_{n,2i})}{n}\om^{*\prime}_i\e_i
\eeq
are unbiased estimators of $\theta$.

\subsection{Transformations of weights and invariance of tests}

We report some theorems on the invariance of the tests under affine transformations. These results can be easily proved and are implicitly used throughout this paper.

\newtheorem{theo}{Theorem} 
\begin{theo}\label{rescale} A test of the form (\ref{testgeno}) does not change its value if all the weights $\w_i$ are rescaled by a common factor $\lambda>0$, that is,
\begin{equation}
\w_i\longrightarrow \lambda \w_i \quad \Rightarrow\quad T_\w \longrightarrow \sign(\lambda) T_\w
\end{equation}\end{theo}
 
Note that the invariance of the tests mean that these transformations define equivalence classes of weights, i.e., sets of different weights that actually correspond to the same test. In particular, this theorem implies that the space of possible tests, in terms of the weights $\w_i$, is not homeomorphic to $\mathbb{R}^{n-2}$ (which would be the subspace of weights in $\mathbb{R}^{n-1}$ that satisfy the linear condition (\ref{sumw})) but to its quotient with respect to the invariance (multiplicative) group $\mathbb{R}^{+}$, that is, the $(n-3)$-dimensional sphere $S^{n-3}=\mathbb{R}^{n-2}/\mathbb{R}^{+}$.

\begin{theo}\label{affine} A test of the form (\ref{testgen}) does not change its value under an affine transformation of parameters $(\lambda,\rho_i)$ on the weights $\om_i$, $\om'_i$ with a common rescaling factor $\lambda>0$, that is,
\begin{equation}
\om_i\longrightarrow \lambda \om_i+\rho_i\ , \  \om'_i\longrightarrow \lambda \om'_i+\rho_i\quad \Rightarrow\quad  T_\w \longrightarrow \sign(\lambda) T_\w
\end{equation}
However, the estimators (\ref{esttheta}) are unbiased only if the rescaling factor satisfies the condition $\lambda=1-\sum_{i=1}^{n-1}\rho_i$.\end{theo}



\subsection{Generalized $D'$ tests for multilocus analysis}

The statistic $D'$ \cite{schaeffer2003molecular}, which is defined as the ratio of Tajima's $D$ versus its minimum value (given a fixed number of segregating sites), has been used in the literature for multilocus analyses \cite{schaeffer2003molecular,schmid2005multilocus,hutter2007distinctly}, arguing that the value of Tajima's $D$  is affected by the length, the sample size and the number of segregating sites of each studied locus and therefore the values of each locus are not directly comparable. 

The contribution of each locus to the heterogeneity is hardly known. Tajima's $D$ is robust to differences in the level of variability (the variance is approximately equal to one)  and also quite robust against  differences in sample size (as is will be shown in the next part), although the quantitative values of Tajima's $D$ for each condition are someway different and therefore the comparison between values is not simple. The proposal of Schaeffer is to use the test
\begin{equation}
D'=\frac{D}{\min(D)_{S=S_{obs}}}
\end{equation}
as a (re)normalized version of Tajima's $D$. $S_{obs}$ is the observed number of segregating sites in the sample.
This proposal can be generalized for all the tests of the form (\ref{testgeno}) as follows:
\begin{equation}
T'_\w=\frac{T_\w}{\min(T_\w)_{S=S_{obs}}}=\frac{\sum_{i=1}^{n-1}i\w_i\xi_i}{\min(j\w_j)S_{obs}}
\end{equation}
This appears to be the natural generalization of $D'$ to general linear tests. 

\section{Sample size independent tests}

\subsection{Scaling of weights with sample size}

In this section we would like to remark that there are conditions that have to be imposed on the weights $\w_i$ or $\om_i,\om'_i$ to ensure that these tests are consistent and meaningful. In fact, the values (and even the number!) of these weights depend explicitly on sample size $n$. Since every conceivable test should be applied to samples of different size, then its definition involves a whole family of weights $\left\{\w_i^{(n)}\right\}$ or $\left\{\om_i^{(n)},\om_i^{\prime (n)}\right\}$ with  ${n=2,3\ldots\infty}$  and to define a test it is necessary to specify how these weights scale with $n$.

\begin{figure}
\begin{center}
\includegraphics[scale=0.5]{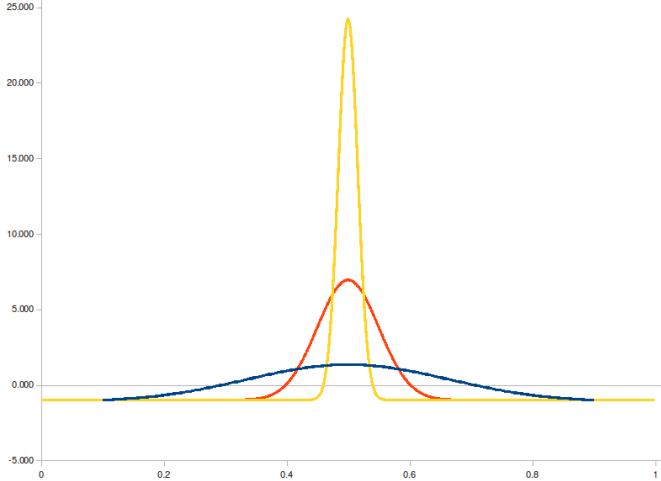} 
\caption{Illustrative example of the dependence of the weight on sample size: weight $\Omega$ as a function of $i/n$ for the test for admixture by Achaz, plotted for different sample size 
$n=10$ (blue), $100$ (red), $1000$ (yellow).}
\label{wf}
\end{center}
\end{figure}

As an example of the weird effects of some choices of scaling, we consider the test for admixture 
of \cite{achaz2009frequency}. The weights of this test are $\om_i={n \choose i}2^{-n}(1-2^{-n+1})^{-1}$ and $\om_i'=1/(n-1)$. Suppose that the population under study shows an excess of alleles of frequency $f$ between 0.3 and 0.4. The average weight of these frequencies, rescaled by the sample size, is 0.5 for $n=10$, but it reduces to -0.75 for $n=100$ and to -1.0 for $n=1000$. These weights are largely different, even in sign, therefore a strong excess of alleles in this range of frequency would show itself as either a positive or a negative value for this test, depending on the sample size! The reason can be understood by noticing that for $n$ large, the binomial can be approximated by a Gaussian function of the allele frequencies $f=i/n$ 
centered in $f=1/2$ and with variance $1/4n$. Therefore this weight function has a strong dependence on $n$ when considered as a
function of $f$ and $n$. The changes of this weight function with sample size are apparent in the plot of Figure \ref{wf}, which shows
the actual function (rescaled by sample size) for $n=10,100,1000$.
In this example it is apparent that the interpretation of the results of this test depends on $n$. This means that the calibration of
the test should be different for each possible sample size. 

The consistency requirement that we propose is that the result of the test should be almost independent on sample size. This requirement is equivalent to a condition on the scaling of the weights $\w_i^{(n)}$ with $n$. Our proposal for a reasonable requirement on this scaling is the following: the relative weight of different frequencies 
in the population should remain approximately constant while varying the size of the sample. This condition ensures that at least for sufficiently large $n$, the average values of the test on samples of different size from the same population should be approximately independent on sample size, i.e. that the test should be consistent.

To determine the scaling, we note that in limit of large $n$, the frequency spectrum approaches a continuum and we can define the weights as functions $\wi(f)$ or $\omi(f),\omi'(f)$ with $f\in (0,1)$ and $\int_0^1df\,\wi(f)=0$, $\int_0^1df\,\omi(f)=\int_0^1df\,\omi'(f)=1$. Since the ratio of the derived allele count and the sample size $i/n$ is an unbiased estimator of the frequency $f$ of the allele in the population (because $E(i)=nf$), a simple scaling that satisfies the above requirement is 
\begin{equation}
\w^{(n)}_i\simeq\wi(i/n) \quad \mathrm{or} \quad \om^{(n)}_i\simeq\omi(i/n)\ ,\ \om^{\prime(n)}_i\simeq\omi'(i/n)\label{scaling}
\end{equation}
as proposed by some of the authors in \cite{ferretti2010optimal}.

In order to have the above approximate scaling while obeying the condition $\sum_{i=1}^{n-1}\w_i=0$, there are two simple consistent forms for the weights:
\begin{equation}
\w^{(n)}_i=\wi\left(\frac{i}{n}\right)-\frac{1}{n-1}\sum_{j=1}^{n-1}\wi\left(\frac{j}{n}\right)\label{scalingnorm}
\end{equation}
where the last term is a (tipically small) correction that enforce centeredness of the test, 
 or
\begin{equation}
\w^{(n)}_i=\om^{(n)}_i-\om^{\prime(n)}_i=\frac{\omi\left(\frac{i}{n}\right)}{\sum_{j=1}^{n-1}\omi\left(\frac{j}{n}\right)}-\frac{\omi'\left(\frac{i}{n}\right)}{\sum_{j=1}^{n-1}\omi'\left(\frac{j}{n}\right)}\label{scalingnorm2}
\end{equation}
where the denominators are normalization factors.

Tipically this second form (\ref{scalingnorm2}) for the scaling is more consistent in practice and it is implicitly assumed for most of the existing tests. However, the above espressions give similar numerical results for most choices of the functions $\wi(f)=\omi(f)-\omi'(f)$. In fact, if $\w(f)$ is a limited and piecewise-continuous function, the difference between (\ref{scalingnorm}) and (\ref{scalingnorm2}) is of order $O({\wi})/n$ (since it is a factor coming from the discretization of the frequencies) and it does not have a relevant impact on the results of the test. Therefore in these cases the two scaling  relations (\ref{scalingnorm}) and (\ref{scalingnorm2}) are practically equivalent. 

Note that all the tests involving the Watterson estimator (that corresponds to $\omi(f)\sim 1/f$) have additional subtleties that are discussed in the next section.

\subsubsection*{Example: Fay and Wu's $H$ test}
This test was proposed in \cite{fay2000hitchhiking} to look for an excess of high-frequency derived alleles as a signal of selection. It can be defined by the weight functions $\omi(f)=2f$ and $\omi'(f)=1$. The weights can be found following equation (\ref{scalingnorm2}). The resulting test is
\begin{equation}
T_{H}=\frac{\frac{2}{n(n-1)}\sum_{i=1}^{n-1}i^2\x_i-\frac{1}{n-1}\sum_{i=1}^{n-1}i\x_i}{\sqrt{\var\left(\frac{2}{n(n-1)}\sum_{j=1}^{n-1}j^2\x_j-\frac{1}{n-1}\sum_{j=1}^{n-1}j\x_j\right)}}
\end{equation}
The scaling defined in equation (\ref{scalingnorm}), with weight function $\wi(f)=\omi(f)-\omi'(f)=2f-1$, gives precisely the same result.

\subsubsection*{Example: $F(r,r')$ tests of Fu \cite{fu1997}}
This large class of test is based on the comparison of two estimators with weights
\begin{equation}
\om_i=\frac{i^{-r}}{\sum_{j=1}^{n-1}j^{-r}}\quad , \quad \om'_i=\frac{i^{-r'}}{\sum_{j=1}^{n-1}j^{-r'}}
\end{equation}
that in the case $r,r'<1$ correspond precisely to the scaling (\ref{scalingnorm2}) suggested above, with weight functions
$\omi(f)=(1-r)f^{-r}$ and $\omi'(f)=(1-r')f^{-r'}$. This can be easily verified by multiplying both the numerator and the denominator of
$\om_i$, $\om'_i$ by a factor $(1-r)/n^{-r}$, $(1-r')/n^{-r'}$ respectively. The test by Fay and Wu corresponds actually to $F(-1,0)$.

The cases with $r\geq 1$ or $r'\geq 1$ involve weight functions with divergent integrals and will be discussed in the next section.

Note that the same weight functions with the scaling (\ref{scalingnorm}) would give rise to a slightly different test with weights
\begin{equation}
\w_i=(1-r)\left(\frac{i}{n}\right)^{-r}- (1-{r'})\left(\frac{i}{n}\right)^{-r'}-\left(\frac{(1-r)\sum_{j=1}^{n-1}j^{-r}}{(n-1)n^{-r}} - \frac{(1-r')\sum_{j=1}^{n-1}j^{-r'}}{(n-1)n^{-r'}} \right) 
\end{equation}
that is not consistent for weights of low frequency alleles, i.e. with $i/n\lesssim n^{2/\max(r,r')}$, and therefore less interesting. 


%



\subsubsection*{Example: test for bottleneck of Achaz \cite{achaz2009frequency}}
This test is another example of a test with an unwanted scaling: 
\begin{equation}
\om_i=\frac{e^{-\alpha i}}{\sum_{j=1}^{n-1}e^{-\alpha j}}\quad , \quad \om'_i=\frac{1}{n-1}
\end{equation}
The weight function for this test is $e^{-\alpha n f}\alpha n /(1-e^{-\alpha n})-1$ that depends strongly on $n$, therefore this test is not consistent in the above sense.

It is easy to build an equivalent test with the correct scaling by choosing the functions $\omi(f)= \beta e^{-\beta f}/(1-e^{-\beta})$, $\omi'(f)=1$. The resulting weights  with the scaling (\ref{scalingnorm2}) are 
\begin{equation}
\om_i=\frac{e^{-\beta i/n}}{\sum_{j=1}^{n-1}e^{-\beta j/n}}=\frac{1-e^{-\beta/n}}{1-e^{-\beta(1-1/n)}}e^{-\beta (i-1)/n}\quad , \quad \om'_i=\frac{1}{n-1}\label{ws1}
\end{equation}
as discussed before. The optimal value reported in \cite{achaz2009frequency} is $\alpha\simeq0.9$ for $n=30$. This value corresponds to $\beta\simeq 27$.

The test can also be implemented by choosing the scaling (\ref{scalingnorm}) and the weight function $\wi(f)=\omi(f)- \omi'(f)=\beta e^{-\beta f}/(1-e^{-\beta})-1$. The resulting weights are
\begin{align}
\w_i=&\frac{\beta e^{-\beta i/n}}{1-e^{-\beta}}-1-\frac{1}{n-1}\left( \frac{\beta(1-e^{-\beta(1-1/n)})}{(1-e^{-\beta})e^{\beta/n}(1-e^{-\beta/n})} -(n-1)\right) =\\ \nonumber
= &  \frac{\beta(1-e^{-\beta(1-1/n)})}{(1-e^{-\beta})e^{\beta/n}(1-e^{-\beta/n})}\cdot \left(\frac{1-e^{-\beta/n}}{1-e^{-\beta(1-1/n)}}e^{-\beta (i-1)/n}-\frac{1}{n-1}\right)
\end{align}
that are equivalent to the weights (\ref{ws1}) up to an irrelevant multiplicative factor (see Theorem \ref{rescale}). Therefore in this case the two choices of scaling give  precisely the same result.

\subsection{Divergent weights}

As discussed above, the two choices of scaling in equation (\ref{scalingnorm}) and (\ref{scalingnorm2}) do not usually bring to sensibly different numerical results. However, there are important choices of  $\wi(f)$ for which this approximate equivalence between (\ref{scalingnorm}) and (\ref{scalingnorm2}) does not hold. These critical cases correspond to functions that diverge as  $1/f$ or faster near $f=0$ (or $f=1$). This divergence is not a real feature of the distribution, because the integral has a natural cutoff at the scale of the inverse population size\footnote{Or more precisely the effective population size $1/N_e$, but this does not affect the above discussion. } $f_{min}=1/N$, but in this case the integral $\int_{1/N}^1 df\,\wi(f)$ has a strong dependence on the cutoff  $1/N$ and therefore the function $\wi(f)$ itself should  depend strongly on $N$ to ensure proper normalization. 

If this dependence is contained in an multiplicative term in front of $\omi(f)$ or $\omi'(f)$ or both,  then the second term in equation (\ref{scalingnorm}) is not a small correction of order $1/n$ as it happens with simple functions $\w(f)$, but rather it represents a relevant correction 
 with a strong dependence on sample size $n$ and population size $N$. 
The denominators in equation (\ref{scalingnorm2}) also show a strong dependence on $n$ (that could not be avoided anyway) but not on $N$, and therefore this second scaling form should be used. The dependence on sample size is as strong as the dependence of the divergent integral from the cutoff\footnote{This can be easily understood by noticing that the sample size $n$ plays the role of the cutoff in the sum over the frequencies that are present in the sample, which is the same role played by the population size $N$ for the whole population. More formally, the denominator in equation (\ref{scalingnorm2}) can be bounded from above and from below by the divergent integral, and therefore the divergence of the denominator as $n\rightarrow\infty$ will be the same as the divergence of the integral as its inverse cutoff (that is, $N$) goes to infinity.}: for functions diverging as $f^{-k}$ with $k\geq 1$, the dependence on $n$ goes as $n^{1-k}$ if $k>1$ or $\log(n)$ for $k=1$. This case always occurs when the test is build by comparing an estimator of $\theta$ with the Watterson estimator, which corresponds to $\omi(f)\sim 1/f$ and therefore has a logarithmic dependence on $n$ given by the usual harmonic factor $a_n=\sum_{j=1}^{n-1}1/j\simeq \log(n)+\gamma+O(1/n)$. A well-known examples of this case is Tajima's $D$ \cite{tajima1989statistical}.

If the dependence of $\wi(f)$ on $N $ is contained in an additive term that does not depend on $f$, it is the correction in (\ref{scalingnorm}) that does not depend on $N$ and therefore the first scaling form is more appropriate. 
We do not know examples of tests of this kind in the literature, even if the test by Zeng \textit{et al.} \cite{zeng2006statistical} can be interpreted also in this way.

\subsubsection*{Example: Tajima's $D$ test}
This is the most known test for neutrality based on the frequency spectrum. It is given by the difference 
between the Tajima estimator 
${\Pi}$  \citep{tajima1983evolutionary} based on the nucleotide pairwise diversity $\Pi$ and the Watterson 
estimator 
${\theta}_W$ \citep{watterson1975number} based on the number $S$ of segregating sites, therefore it can be defined 
by the weight functions $\omi(f)=2(1-f)$ for $\Pi$ and $\omi'(f)=1/f\log(N)$ for the Watterson estimator. The latter 
function has an integral that diverges logarithmically near $f=0$, and the corresponding dependence on $N$ is contained in the factor $1/\log(N)$ that multiplies $\omi'(f)$, therefore the scaling (\ref{scalingnorm2}) should be used. The result is the usual test 
\begin{equation}
T_D=\frac{\sum_{i=1}^{n-1}\frac{2i(n-i)}{n(n-1)}\x_i-S/a_n}{\sqrt{\var\left(\sum_{j=1}^{n-1}\frac{2j(n-j)}{n(n-1)}\x_j-S/a_n\right)}}=\frac{\Pi-S/a_n}{\sqrt{\var\left(\Pi-S/a_n\right)}}
\end{equation}

\subsubsection*{Example: test of Zeng \textit{et al.} \cite{zeng2006statistical}}
This test was proposed to look for an excess of high-frequency derived alleles compared to low-frequency alleles. 
It is defined by the weight functions $\omi(f)=1$ and $\omi'(f)=1/f\log(N)$, the latter corresponding to the Watterson estimator. 
Proceeding as in the above example, the result is
\begin{equation}
T_{E}=\frac{\sum_{i=1}^{n-1}\frac{i}{(n-1)}\x_i-S/a_n}{\sqrt{\var\left(\sum_{j=1}^{n-1}\frac{j}{(n-1)}\x_j-S/a_n\right)}}
\end{equation}
Note that exceptionally the scaling of this test can also be defined by (\ref{scalingnorm}), without modifying the result. This is a consequence 
of the two equivalent forms for the weight function, $\wi(f)=1-1/f\log(N)$ or $\wi(f)=\log(N)-1/f$.

\subsection{Weights of singletons}

The above scaling (\ref{scaling}) is valid in principle for all weights. However in practice there is an important exception, that is, the weight $\w_1$ of singletons. This is due to the fact that for $n\ll N$,  
the number of derived singletons $\x_1$ is the only estimator that is affected by very rare derived alleles (and often by sequencing errors, see \cite{achaz2008testing}). More precisely, $\x_1$  is actually the only estimator sensitive to the deviations from neutrality in alleles of frequency $1/N<f<1/n$, which represent a vast majority of the SNPs in the population and can contain interesting biological information. Therefore, if the contribution of these alleles is relevant for the test, we can enhance (or reduce) the weight $\w_1$ by adding a factor $\w_{ds}$. 

In the approach detailed in the previous sections, this additional contribution to $\w_1$ is 
needed to take into account a contribution $\Delta\w(f)$ to $\w(f)$ of the form $\Delta\w(f)=\w_{ds}I(f<\phi)/\phi$ with $\phi \ll 1$. 
As far as the maximum sample size never exceeds in practice $n_{max}\ll 2/\phi$, this  function  weights positively only alleles that appear as singletons.

Similarly, $\om_1$ and $\om'_1$ can be enhanced by $\om_{ds}$, $\om'_{ds}$ that correspond to contributions $\Delta\om(f)=\om_{ds}I(f<\phi)/\phi$, $\Delta\om'(f)=\om'_{ds}I(f<\phi)/\phi$. The test of Fu and Li \cite{fu1993statistical} fall into this case.

A similar argument applies also to the weights of the number of ancestral singletons, that is, $\w_{n-1}$, $\om_{n-1}$, $\om'_{n-1}$ that can be enhanced by factors $\w_{as}$, $\om_{as}$ and $\om'_{as}$ respectively. However this case is more rare, the only interesting example being the tests of Achaz \cite{achaz2008testing} that avoid sequencing errors by neglecting both derived and ancestral singletons.


Summarizing the results up to this section, a test $T_\w$ is completely defined by a function $\wi(f)$ and two parameters $\w_{ds}$, $\w_{as}$ (that could depend on $n$) satisfying the conditions
\begin{equation}
\w_{ds}+\w_{as}+\int_0^1df\,\wi(f)=0
\end{equation}
and determining the weights through the formula:  
\begin{equation}
\w^{(n)}_i=\wi\left(\frac{i}{n}\right)+\w_{ds}\delta_{i,1}+\w_{as}\delta_{i,n-1}-\frac{1}{n-1}\left(\w_{ds}+\w_{as}+\sum_{j=1}^{n-1}\wi\left(\frac{j}{n}\right)\right)\label{scalingsingl}
\end{equation}
or by a pair of functions $\omi(f),\omi'(f)$ and parameters $\om_{ds}$, $\om'_{ds}$, $\om_{as}$, $\om'_{as}$ satisfying
\begin{equation}
\om_{ds}+\om_{as}+\int_0^1df\,\omi(f)=\om'_{ds}+\om'_{as}+\int_0^1df\,\omi'(f)=1
\end{equation}
and resulting in this formula for the scaling of the weights:
\begin{equation}
\w^{(n)}_i=\frac{\om_{ds}\delta_{i,1}+\om_{as}\delta_{i,n-1}+\omi\left(\frac{i}{n}\right)}{\om_{ds}+\om_{as}+\sum_{j=1}^{n-1}\omi\left(\frac{j}{n}\right)}-\frac{\om'_{ds}\delta_{i,1}+\om'_{as}\delta_{i,n-1}+\omi'\left(\frac{i}{n}\right)}{\om'_{ds}+\om'_{as}+\sum_{j=1}^{n-1}\omi'\left(\frac{j}{n}\right)}
\end{equation}

As showed in the examples above and below, most of the tests in the literature have this general scaling, with the only exceptions of the ones contained in \cite{achaz2009frequency} that are not consistent in the above sense. 

\subsubsection*{Example: Fu and Li's $F$ test}
This test looks for an excess of very rare derived alleles as a possible signature of negative selection \cite{fu1993statistical}. The only nonzero weights are $\om_{ds}=1$ and  $\omi'(f)=1/f\log(N)$, while $\omi(f)=\om'_{ds}=\om_{as}=\om'_{as}=0$. The resulting test is
\begin{equation}
T_{F}=\frac{\x_1-S/a_n}{\sqrt{\var\left(\x_1-S/a_n\right)}}
\end{equation}
Note that this test has both singleton weights and a divergent weight function.

\subsubsection*{Example: error-corrected tests of Achaz \cite{achaz2008testing}}
This class of tests is an attempt to correct for sequencing errors and biases in the data by removing the alleles where most of the problems manifest themselves, i.e. singletons (both ancestral and derived). With a slight generalization of the proposal in \cite{achaz2008testing}, the weights of the singletons are chosen in such a way to cancel precisely the contributions of the weight functions:
\begin{equation}
\w_{ds}=-\wi\left(\frac{1}{n}\right),\w_{as}=-\wi\left(1-\frac{1}{n}\right)
\end{equation}
or
\begin{equation}
\om_{ds}=-\omi\left(\frac{1}{n}\right),\om_{as}=-\omi\left(1-\frac{1}{n}\right),\om'_{ds}=-\omi'\left(\frac{1}{n}\right),\om'_{as}=-\omi'\left(1-\frac{1}{n}\right)
\end{equation} 
therefore the final weights of derived or ancestral singletons are zero. These corrections can be applied in principle to any weight function.

\subsection{Scaling of weights in tests without an outgroup}

The above arguments can be repeated in a straightforward way for the tests $T^*_\w$ based on the folded spectrum $\e_i$. The only relevant difference is that the frequency $f$ of the minor allele in the population is always less than 50\%, that is, $f\in (0,1/2]$. For consistency with the unfolded case, the weight $\e_{n/2}$ is reduced by a factor $2$. Moreover, the additional parameters related to the weights of singletons cannot distinguish between ancestral and derived alleles and therefore reduce to $\w^*_s$, $\om^*_s$, $\om^{*\prime}_s$. These parameter, together with the functions $\wi^*(f)$, $\omi^*(f)$ and $\omi^{*\prime}(f)$, should satisfy the conditions  
\begin{equation}
\w^*_{s}+\int_0^{1/2}df\,\wi^*(f)=0 \quad , \quad \om^*_{s}+\int_0^{1/2}df\,\omi^*(f)=\om^{*\prime}_{s}+\int_0^{1/2}df\,\omi^{*\prime}(f)=1
\end{equation}

The formulae that determine the scaling of the weights are:  
\begin{equation}
\w^{*(n)}_i=\frac{1}{1+\delta_{n,2i}}\wi^*\left(\frac{i}{n}\right)+\w^*_{s}\delta_{i,1}-\frac{1}{\fln}\left(\w^*_{s}+\sum_{j=1}^{\fln}\frac{1}{1+\delta_{n,2j}}\wi^*\left(\frac{j}{n}\right)\right)
\end{equation}
\begin{equation}
\w^{*(n)}_i=\frac{\om^*_{s}\delta_{i,1}+\omi^*\left(\frac{i}{n}\right)/(1+\delta_{n,2i})}{\om^*_{s}+\sum_{j=1}^{\fln}\omi^*\left(\frac{j}{n}\right)/(1+\delta_{n,2j})}-\frac{\om^{*\prime}_{s}\delta_{i,1}+\omi^{*\prime}\left(\frac{i}{n}\right)/(1+\delta_{n,2i})}{\om^{*\prime}_{s}+\sum_{j=1}^{\fln}\omi^{*\prime}\left(\frac{j}{n}\right)/(1+\delta_{n,2j})}
\end{equation}

The weights of the folded versions of Tajima's $D$ and Fu and Li's $F^*$ test follow this scaling. The nonzero weight functions are $\omi^*(f)=1$, $\omi^{*\prime}(f)=1/(\log(N)f(1-f))$ for Tajima's $D$ and $\om^*_{s}=1$, $\omi^{*\prime}(f)=1/(\log(N)f(1-f))$ for the test of Fu and Li.

\subsection{Alternative choices of scaling}
The choice of scaling discussed in the previous sections represents a quite simple and effective way to fix the dependence on $n$ of a newly devised test. However, other choices are possible whose weights differ from the above ones for small $n$. The reason is that for $n$ not too large, both the variance of order $f(1-f)/n\simeq i(n-i)/n^3$ in the estimation of the frequency $f=i/n$ and the related uncertaincy about how the frequencies are actually weighted in the test become important. This uncertaincy originates from the (binomial) sampling of individuals from the population and there is some degree of arbitrariness in deciding how to account for it. Moreover, tests that take it into account could be not consistent in the above sense.

A possible choice of scaling that uses the binomial sampling is the following: considering $\omi(f)$, $\omi'(f)$ as frequency distributions, the weights $\om_i$, $\om'_i$ are assigned from $\omi(f)$, $\omi'(f)$ through the same binomial sampling that is done for allele spectra, that is,
\begin{equation}  
\om_i=\frac{\int_0^1 df\,{n \choose i}f^i(1-f)^{n-i}\,\omi(f)}{\int_0^1 df\,(1-f^n-(1-f)^n)\,\omi(f)} \label{strangescal1}
\end{equation}
\begin{equation}
\omi'_i=\frac{\int_0^1 df\,{n \choose i}f^i(1-f)^{n-i}\,\omi'(f)}{\int_0^1 df\,(1-f^n-(1-f)^n)\,\omi'(f)}\label{strangescal2}
\end{equation}

A simple example of this scaling (but with an highly divergent weight function) is given by the test for admixture \cite{achaz2009frequency} discussed before. Optimal tests also follow this scaling. 

\subsubsection*{Example: test for admixture of Achaz \cite{achaz2009frequency}}
This test is apparently not consistent and it does not follow the scaling (\ref{scaling}). 
However it follows another scaling related to the allele sampling. To understand this, consider the weight functions $\omi(f)=\delta(f-1/2),\omi'(f)=1$ where $\delta(f-1/2)$ is a Dirac delta function\footnote{The Dirac delta $\delta(f-a)$ is a function whose value is $0$ if $f\neq a$ and $+\infty$ if $f=a$. The integral $\int \delta(f-a)g(f)df$ is $g(a)$ if $a$ is inside the range of integration and $0$ otherwise. Actually this function is not a mathematical function, but a distribution, i.e. an element of a dual space of regular functions.} centered in $1/2$. If we scale the weights according to (\ref{strangescal1}),(\ref{strangescal2}), that is,
\begin{equation}  
\om_i=\frac{\int_0^1 df\,{n \choose i}f^i(1-f)^{n-i}\,\omi(f)}{\int_0^1 df\,(1-f^n-(1-f)^n)\,\omi(f)}=\frac{{n \choose i}2^{-n}}{1-2^{-n+1}} 
\end{equation}
\begin{equation}
\omi'_i=\frac{\int_0^1 df\,{n \choose i}f^i(1-f)^{n-i}\,\omi'(f)}{\int_0^1 df\,(1-f^n-(1-f)^n)\,\omi'(f)}=\frac{1}{n-1}
\end{equation}
then the corresponding test is precisely the one proposed by Achaz. Note that the strong dependence of the test from sample size does not come only from the choice of scaling, but also from the weight function chosen, that is highly divergent.
 
\section{Linear optimal tests}

\subsection{On the existence of generic tests}
An interesting question on the way to build good linear tests is the following: do there exist generic tests?
A completely generic test for neutrality should be able to detect any deviation from the spectrum of the null model that is sufficiently large. Unfortunately, these tests do not exist. In fact, for every  test defined by a set of weights $\w_i$ it is possible to find a spectrum $\x_i=\alpha/{ia_n}+(1-\alpha)\Delta_i$ 
that is maximally different from the standard spectrum at least in a range of frequencies and is nevertheless undetectable by the test because its average value on this spectrum is zero. This is expressed in a more formal way in the following theorem, which shows that even the complete lack of alleles in some range of frequencies could not be always detected.

\begin{theo}\label{theo1} For every set of $n$ real weights $\w_i$ with $\sum_i\w_i=0$, there is a set of $n$ real numbers $\Delta_i\neq const/i$ and a parameter  $\alpha\in[0,1]$ that satisfy the conditions
\begin{equation}
\sum_ii\w_i\Delta_i=0\quad,\quad\min_{i\in [1,n-1]}\left(\alpha\frac{1}{ia_n}+(1-\alpha)\Delta_i\right)=0 
\end{equation}\end{theo}
\newproof{prove}{Proof}

The above limitation is not a consequence of the small sample size. This can be seen for example in the framework of the scaling theory discussed in this paper. In fact, for large sample size, the weights can be approximated by a weight function $\wi(f)$. In this context it is possible to prove the next theorem, that is a continuous equivalent of the previous one.
\begin{theo}\label{theo2} For every piece-wise continuous weight function $\wi(f)\in L^1_{[1/N,1]}$ such that $\int_{1/N}^1\wi(f)df=0$, there is a smooth function $\Delta(f)\neq const/f$ and a parameter $\alpha\in[0,1]$ that satisfy the conditions
\begin{equation}
\int_{1/N}^1df\, f \wi(f)\Delta(f)=0\quad,\quad\inf_{f\in[0,1]}\left(\alpha\frac{1}{f\log(N)}+(1-\alpha)\Delta(f)\right)=0
\end{equation}
 \end{theo}

Note that in principle this problem can be solved using multiple tests. In fact multiple tests should be able to detect any strong deviation from the null spectrum, provided that the number of these tests is large enough, as can be seen from the following theorem.
\begin{theo} \label{theo3} Given at least $n-2$ linearly independent sets of $n-1$ real weights $\w_i$ with $\sum_i\w_i=0$, it is not possible to find a set of real numbers $\Delta_i\neq const/i$ such that $\sum_ii\w_i\Delta_i=0$.
\end{theo}

This last theorem is only a formal result and the requirement of $n-2$ independent tests is too strong. In practice a small (but good) set of tests can detect most of the reasonable and interesting deviations for realistic spectra.

The above theorems can be extended to the folded spectrum. In this section and the next ones, we will consider only tests based on the unfolded spectrum. The generalization of the discussion to the folded spectrum is usually straightforward after substituting $\xi_i$ ($i=1\ldots n-1$) with $\eta_i$ ($i=1\ldots\fln$).

\subsection{Optimal tests and their geometric structure}

From the theorems of the previous section, it is clear that a single test cannot detect all the possible deviations occurring in complicated evolutionary scenarios. However it is still possible to optimize neutrality tests of for a specific alternative evolutionary scenario. A simple optimality condition has been proposed by some of the authors in \cite{ferretti2010optimal} in order to maximize the power of the test to detect a fixed alternative scenario. If the null spectrum is ${E}(\xi_i)=\theta L\xz_i$ and the expected spectrum of the alternative scenario is $\mathcal{E}(\xi_i)=\theta L\xb_i$, the condition for optimal tests is the maximization of the average result of the test under the alternative scenario:
\begin{equation}
\mathcal{E}(T_\w)=\frac{\sum_{i=1}^{n-1}\w_i\theta L\xb_i/\xz_i}{\sqrt{\var\left( \sum_{j=1}^{n-1}\w_j\x_j/\xz_j\right)}}
\end{equation} 
This condition is based on the observation that the tests have mean zero and variance $1$, therefore if the distributions of the results of the tests are similar, the maximization of the average value of the test should correspond to the maximization of the average power of the test. It is also possible to maximize directly the power of the test, taking into account the different distribution of the results under the null and the alternative model; this  possibility will be pursued in section \ref{maxpow}.

\begin{center}
\begin{figure}
 \includegraphics[scale=0.25]{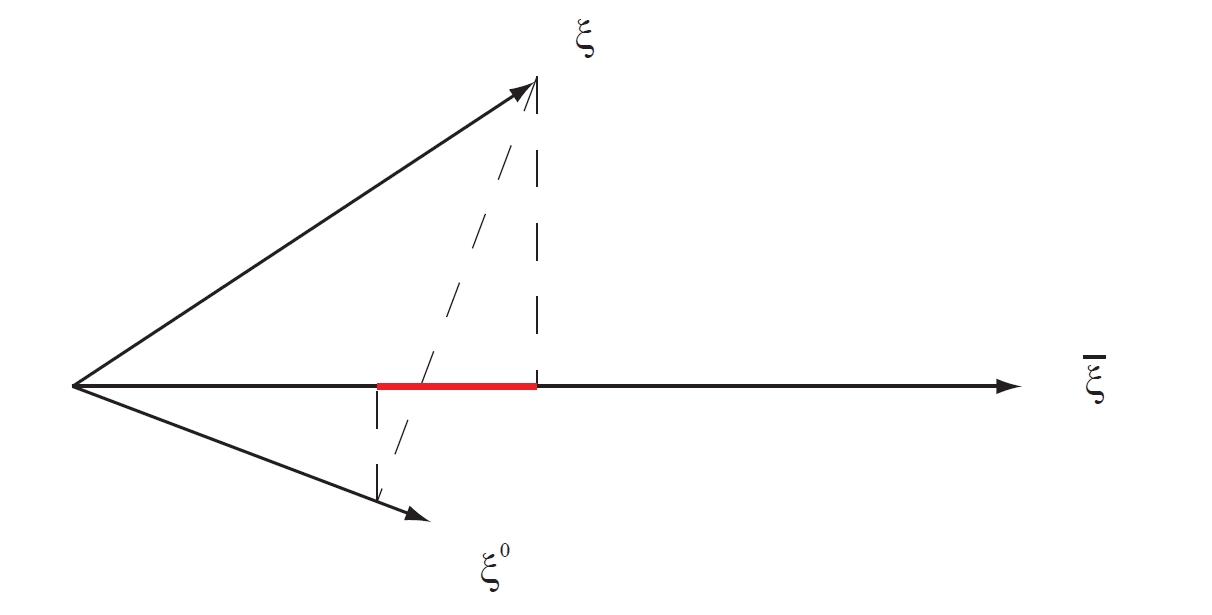} 
\caption{Geometrical representation of the numerator of the optimal test $\topt$ in (\ref{testscal}). The length of the red line segment corresponds to the value of the numerator.}
\label{vectors}
\end{figure}
\end{center}

Interestingly, optimal tests show a geometric structure which becomes apparent after defining the scalar product between spectra:
\begin{equation}
\left\langle\langle\xi',\xi'' \right\rangle\rangle \equiv \sum_{i,j} \xi'_i c_{ij}^{-1} \xi''_j\label{scalprod}
\end{equation}
where $c_{ij}^{-1}$ is the inverse of the covariance matrix $\cov(\xi_i,\xi_j)$. Since $\cov(\xi_i,\xi_j)$ is symmetric and positive, its inverse is also symmetric and positive, i.e. it is a positive bilinear form, therefore the above expression defines a scalar product. Then the optimal test for an alternative spectrum $\xb$ can be written in the elegant form\footnote{We do not provide a proof of this expression here because it can be easily obtained as a special case of the general formula (\ref{opts}) that we will discuss later in the context of nonlinear tests. A direct proof of this result can be found in \cite{ferretti2010optimal} after substituting the scalar products with the definition (\ref{scalprod}).}
\begin{equation}
\topt=\frac{\left\langle\langle\xi,\xb\right\rangle\rangle-\left\langle\langle\xi,\xz\right\rangle\rangle{\left\langle\langle\xz,\xb\right\rangle\rangle}/{\left\langle\langle\xz,\xz\right\rangle\rangle}}{\sqrt{
\left\langle\langle\xb,\xb\right\rangle\rangle-\left\langle\langle\xz,\xb\right\rangle\rangle^2/\left\langle\langle\xz,\xz\right\rangle\rangle
}} \label{testscal}
\end{equation}
The numerator of the test is actually the matrix element between $\xb$ and $\xi$ of the linear operator $1-P_{\xz}$, where $P_{\xz}$ is
the projection operator along $\xz$. In other words, it is proportional to the difference between the length of the projection of $\xi$
on $\xb$ and the length of the projection on $\xb$ of the spectrum obtained by the projection of $\xi$ on $\xz$, as illustrated in
Figure \ref{vectors}.
From this geometrical interpretation it is clear that if the spectrum $\xi$ corresponds to the null spectrum $\xz$, then the two projections are equal and the result of the test is zero. On the other side, if the spectrum is the alternative spectrum $\xb$, then the value of the test is 
\begin{equation}
\topt^{(max)}=\theta L\sqrt{\left\langle\langle\xb,\xb\right\rangle\rangle-\frac{\left\langle\langle\xz,\xb\right\rangle\rangle^2}{\left\langle\langle\xz,\xz\right\rangle\rangle}}
\end{equation}
which is the maximum value over all possible tests in the alternative scenario. The same expression, but with a minus sign, corresponds to the minimum value.

The denominator of the test is the square root of the matrix element of the linear operator $1-P_{\xz}$ between $\xb$ and itself. Note that both the numerator and the denominator of the test do not change by adding any (possibly negative) multiple of $\xz$ to $\xb$, because $\xz$ lies in the kernel of $1-P_{\xz}$. This means that optimal tests depend only on the expected deviations from the null spectrum in the alternative scenario. The result of the test is maximum when the deviations of the data from the null spectrum correspond exactly to the expected ones, and it is minimum when they are opposite to the expected ones.  


\subsection{Scaling of optimal tests}


Optimal tests have weights proportional to the expected allele distribution $\om_i=\xb_i/\sum_{j=1}^{n-1}\xb_j$ and to the null allele distribution $\om'_i=\xz_i/\sum_{j=1}^{n-1}\xz_j$. 
Therefore the weights of an optimal test follow the same scaling with sample size as the allele distributions. Denoting by $\xb(f)$ and $\xz(f)$ the spectra of expected and null allele frequencies in the whole population, the spectra for the sample are obtained by binomial sampling 
\beq
\xi_i=\int_{1/N_e}^1 df\, P_{bin}(i;n,f) \xi(f)\qquad P_{bin}(i;n,f)={n \choose i} f^i (1-f)^{n-i}
\eeq
from the spectra $\xi(f)=\xb(f)$ and $\xi(f)=\xz(f)$ respectively. Therefore the scaling of the allele distributions is 
\begin{equation}  
\frac{\xb_i}{\sum_{j=1}^{n-1}\xb_j}=\frac{ \int_{1/N_e}^1 df\,{n \choose i}f^i(1-f)^{n-i}\,\xb(f) }{ \int_{1/N_e}^1 df\,(1-f^n-(1-f)^n)\,\xb(f) }
\end{equation}
\begin{equation}  
\frac{\xz_i}{\sum_{j=1}^{n-1}\xz_j}=\frac{ \int_{1/N_e}^1 df\,{n \choose i}f^i(1-f)^{n-i}\,\xz(f) }{ \int_{1/N_e}^1 df\,(1-f^n-(1-f)^n)\,\xz(f) }
\end{equation}
which is precisely the scaling (\ref{strangescal1}),(\ref{strangescal2}) with weight functions $\omi(f)\propto \xb(f)$ and $\omi'(f)\propto \xz(f)$.
This scaling does not correspond to the scaling (\ref{scaling}) suggested in this paper, but it takes into account the sampling process in a straightforward way, being based on the expected and null allele distributions for the sample and therefore immediately related to the binomial sampling of alleles from the population. 

Note that these weights actually follow the scaling (\ref{scaling}) for large $n$, with the same weight functions $\omi(f)\propto\xb(f)$ and $\omi'(f)\propto \xz(f)$. This agrees with the fact that when the sample size is large enough, the variance of the sampling process can be safely ignored and all reasonable choices of scaling are equivalent to our proposal (\ref{scaling}).

\subsection{Generalizing $D'$ for optimal test }
The $D'$ statistics of \cite{schaeffer2003molecular} can be generalized for optimal tests as it was done for general linear tests. In particular, for a fixed number of segregating sites $S_{obs}$, the generalization for optimal tests in the approximation of unlinked sites and $\theta\ll 1$ is
\begin{equation}
\topt^\prime=\frac{\sum_{i=1}^{n-1}\w_i\xi_i/\xz_i}{\min_{j}
(\w_j/\xz_j){S_{obs}}}=\frac{\sum_{i=1}^{n-1}(\xi_i/S_{obs})(\xb_i/\sum_{j=1}^{n-1}\xb_j)/(\xz_i/\sum_{j=1}^{n-1}\xz_j)-1}{ 
\min_{k}\left((\xb_k/\sum_{l=1}^{n-1}\xb_l)/(\xz_k/\sum_{l=1}^{n-1}\xz_l)\right)-{1}}
\end{equation}
which has the interesting property of depending only on the allele frequency distributions. 

However in the case of optimal tests there is another possibility, namely to define a $\bar{D}'$ test as the ratio of the optimal test and of its average minimum, assuming that the spectrum corresponds to the average spectrum of the actual scenario. This would correspond to the form
\begin{equation}
\bar{T}^\prime_O=-\frac{\left\langle\langle\xi,\xb\right\rangle\rangle-\left\langle\langle\xi,\xz\right\rangle\rangle{\left\langle\langle\xz,\xb\right\rangle\rangle}/{\left\langle\langle\xz,\xz\right\rangle\rangle}}{S_{obs}/a_n\sqrt{
\left(
\left\langle\langle\xb,\xb\right\rangle\rangle-\left\langle\langle\xz,\xb\right\rangle\rangle^2/\left\langle\langle\xz,\xz\right\rangle\rangle
\right)\left(
\left\langle\langle\xi,\xi\right\rangle\rangle-\left\langle\langle\xz,\xi\right\rangle\rangle^2/\left\langle\langle\xz,\xz\right\rangle\rangle
\right)
}} \label{testd2}
\end{equation}
which has the interesting property of being symmetric with respect to the actual spectrum $\xi$ and the expected spectrum $\xb$.


\subsection{Linear tests with maximum power}\label{maxpow}

The condition for optimal tests is the maximization of $\mathcal{E}(T_\w)$ under the alternative scenario. However, a better approach would by the maximization of the power of the test to reject the neutral model in the alternative scenario, given a choice of significance level $\alpha$. 
This approach require the knowledge of the form of the probability distributions $p(T_\Omega=t|H_0)$, $p(T_\Omega=t|H_1)$ where $H_0$ and $H_1$ are the null and alternative model, or equivalently of all the moments of the spectrum ${E}(\xi_i\xi_j\xi_k\ldots)$ and $\mathcal{E}(\xi_i\xi_j\xi_k\ldots)$. 

Since this information is usually not available in analytic form and hard to obtain computationally, we limit to the case where the distribution can be well approximated by a Gaussian both for the null and for the expected model. Then the only information neeeded are the spectra $\mu_i=E(\xi_i)$, $\bar{\mu}_i=\mathcal{E}(\xi_i)$ and their covariance matrices $c_{ij}=E(\xi_i\xi_j)-E(\xi_i)E(\xi_j)$, $\bar{c}_{ij}=\mathcal{E}(\xi_i\xi_j)-\mathcal{E}(\xi_i)\mathcal{E}(\xi_j)$. 

We expect that both in this approximation and in the general case, the tests with maximum power will depend on the significance level chosen, therefore limiting the interest of these test and the possibilities of comparison between results of the test on samples from different experiments. 

We call $\tau=\mathrm{erf}^{-1}(1-2\alpha)$ the $z$-value corresponding to the critical $p$-value $\alpha$. In the Gaussian approximation, the power is given by the following expression
\beq
\mathrm{Power}=\frac{1}{2}\left[ 1+\mathrm{erf}\left(\frac{\sum_j\bar{\mu}_j\Omega_j-\tau\sqrt{\sum_{j,k}c_{jk}\Omega_j\Omega_k}}{\sqrt{\sum_{j,k}\bar{c}_{jk}\Omega_j\Omega_k}}\right) \right]
\eeq
then its maximization is equivalent to the maximization of 
\beq
\frac{\sum_j\bar{\mu}_j\Omega_j-\tau\sqrt{\sum_{j,k}c_{jk}\Omega_j\Omega_k}}{\sqrt{\sum_{j,k}\bar{c}_{jk}\Omega_j\Omega_k}}\label{eqmaxpow}
\eeq 
In the general case, the weights corresponding to the maximum depend explicitly on $\tau$ and therefore on $\alpha$. This dependence is expected but unwanted, since the interpretation of the test depends explicitly on the critical $p$-value chosen. Obtaining explicit solutions for the weights is complicated and will not be discussed here. 

There is only one case with weights independent on $\tau$, that is the case of $\bar{c}_{ij}$ (approximately) proportional to ${c}_{ij}$. In this case the maximization of the power of the test is (approximately) equivalent to the maximization of the average result of the test, which is precisely our condition for optimal tests.

There is also a regime of values of $\alpha$ such that the weights corresponding to maximum power are independent on $\alpha$, that is, the regime $\tau(\alpha)\gg 1$. In this case the power is an increasing function of ${\sum_{j,k}\bar{c}_{jk}\Omega_j\Omega_k}/{\sum_{l,m}{c}_{lm}\Omega_l\Omega_m}$ and the weights are simply given by the null eigenvector (or linear combination of null eigenvectors) of the matrix $\bar{c}_{ij}-\chi c_{ij}$, where $\chi$ is uniquely defined by the requirement that $\bar{c}_{ij}-\chi c_{ij}$ be a negative semidefinite matrix with at least a null eigenvalue. However, this regime is uninteresting because such small significance levels are practically useless (if $\tau\sim 10$, the corresponding critical $p$-value is $\alpha\sim 10^{-20}$).

In our opinion, the maximization of (\ref{eqmaxpow}) is not interesting in practice because of the dependence on $\alpha$ and of the
complicated form of the corresponding weights. Optimal tests represent a good compromise between high power, simplicity and easiness to
interpret and compare the results. However, it could be possible to build interesting tests with higher power by selecting a linear
combinations of the weights of the two $\alpha$-independent tests discussed above, that is, optimal tests and tests that maximize the
alternative/null variance ratio ${\sum_{j,k}\bar{c}_{jk}\Omega_j\Omega_k}/{\sum_{l,m}{c}_{lm}\Omega_l\Omega_m}$.

\section{Beyond linear neutrality tests}

\subsection{Quadratic and nonlinear tests}

Almost all the neutrality tests proposed in the literature are linear in the spectrum $\xi_i$. As far as we know, there is only one exception, 
namely the $G_\xi$ test of Fu \cite{fu1996new}. This test is a quadratic polynomial 
reminescent of
Hotelling's $t^2$ statistics for the different components of the spectra:
\beq
G=\sum_{i,j=1}^{n-1}c_{ij}^{-1}(\xi_i-\theta L \xz_i) (\xi_j-\theta L \xz_j)\label{fuquad}
\eeq
where $c_{ij}^{-1}$ is the inverse of the covariance matrix $\cov(\xi_i,\xi_j)$. Actually the test proposed by Fu is an approximation to this test with a different normalization, namely 
\beq
G_\xi=\frac{1}{n-1}\sum_{i=1}^{n-1}\frac{(\xi_i-\theta L \xz_i)^2}{\var{(\xi_i)}}\label{fug}
\eeq
In this approximation the off diagonal terms in the covariance can be neglected \cite{fu1995statistical,fu1996new}. For large samples, the distribution of the results of the test $G$ tends to a $\chi^2$ distribution with $n-1$ degrees of freedom.  

Fu's approach cannot be extended to general quadratic or higher order tests, because the distribution of the results of the test would be generally unknown and not positive definite. For this reason we propose to rescale the tests to have zero mean and variance $1$. With this normalization, we expect that the distribution would asymptotically converge to a Gaussian $N(0,1)$ for all tests. 
As an example, the (re)normalized version of Fu's test would be
\beq
T_G=\frac{\sum_{i,j=1}^{n-1}c_{ij}^{-1}(\xi_i-\theta L \xz_i) (\xi_j-\theta L \xz_j)-(n-1)}{\sqrt{\var\left(\sum_{i,j=1}^{n-1}c_{ij}^{-1}(\xi_i-\theta L \xz_i) (\xi_j-\theta L \xz_j)-(n-1)\right)}}
\eeq
Since the only difference between this test and the original one is the normalization and a shift by a constant factor $n-1$, the power of the test is the same. 

Now we present a systematic discussion of nonlinear tests that are generic polynomials (or eventually power series) in the spectrum $\xi_i$. All the tests are rescaled to be centered (i.e., to have zero mean) and have variance $1$. We denote by $\mu_{ijk\ldots}$
the moments of the spectrum under the null model, that is, $\mu_{ijk\ldots}=E(\xi_i\xi_j\xi_k\ldots)$. With this definition, $\mu_i=\theta L \xz_i$. Note that all these moments depend on $\theta$. In the approximation of unlinked (independent) sites and small $\theta$, 
the second moments are equal to $\mu_{ij}=\theta L \xz_i\delta_{ij}+\theta^2 L^2 \xz_i\xz_j$.

The weights of general nonlinear tests can depend explicitly on $\theta$, as seen in the previous example. To compute the values of the tests, the (unknown) parameter $\theta$ is substituted with an estimator $\hat{\theta}$. Unlike the linear case, in this case there are two different classes of tests, related to the dependence on $\hat{\theta}$ of the centeredness:
\emph{strongly centered} and  \emph{weakly centered} tests. 

Strongly centered tests are tests that are always centered for any value of $\hat{\theta}$, even if it is different from the actual value of $\theta$. The general form for strongly centered tests is
\begin{equation}
T_\w=\frac{\sum_{i=1}^{n-1}\w_i^{(1)}\xi_i+\sum_{i,j=1}^{n-1}\w_{ij}^{(2)}\xi_i\xi_j+\sum_{i,j,k=1}^{n-1}\w_{ijk}^{(3)}\xi_i\xi_j\xi_k+\cdots}{\sqrt{\var\left( {\sum_{i=1}^{n-1}\w_i^{(1)}\xi_i+\sum_{i,j=1}^{n-1}\w_{ij}^{(2)}\xi_i\xi_j+\sum_{i,j,k=1}^{n-1}\w_{ijk}^{(3)}\xi_i\xi_j\xi_k+\cdots}\right)} }
\end{equation}
with the real symmetric weights $\w_{ijk\ldots}^{(n)}$ satisfying the set of conditions 
\begin{equation}
0=\sum_{i=1}^{n-1}\w_i^{(1)}\mu_i^{(m)}+\sum_{i,j=1}^{n-1}\w_{ij}^{(2)}\mu_{ij}^{(m)}+\ldots\quad,\quad m=1,2,3\ldots
\end{equation}
where we denote by 
$\mu_{ijk\ldots}^{(p)}$ the $p$-th term of the Taylor expansion with respect to $\theta L$ of 
$\mu_{ijk\ldots}$
\footnote{In other words, 
$\mu_{ijk\ldots}=\sum_p\theta^pL^p\mu_{ijk\ldots}^{(p)}$, where the coefficients 
$\mu_{ijk\ldots}^{(p)}$ are independent on $\theta$.}.
The sum can be limited to polynomials of some finite order in $\xi_i$ or it can be a (convergent) power series.
If we introduce the notation $\mathbf{I}=ijk\ldots$ to denote a group of $n_\mathbf{I}$ indices, we can rewrite the test in the simpler form
\begin{equation}
T_\w=\frac{\sum_{\mathbf{I}}\w_\mathbf{I}^{(n_\mathbf{I})}(\xi\ldots\xi)_{\mathbf{I}}}{\sqrt{\var\left( {\sum_{\mathbf{I}}\w_\mathbf{I}^{(n_\mathbf{I})}(\xi\ldots\xi)_{\mathbf{I}}} \right)}} \label{formstrong}
\end{equation}
with the conditions
\begin{equation}
0=\sum_{\mathbf{I}}
\w_\mathbf{I}^{(n_\mathbf{I})}\mu_\mathbf{I}^{(m)}\quad,\quad m=1,2,3\ldots\label{condstrong}
\end{equation}
If we constrain these tests to be first order polynomials in $\xi_i$, we recover the linear case with $\w^{(1)}_i=\w_i/\xz_i$. Note that linear tests are always strongly centered. In fact in the infinite site model the spectrum is always proportional to $\theta$, which consequently factorizes out by linearity and therefore has no effect on the centeredness.

Weakly centered tests are tests that are centered but not strongly centered, i.e., they are centered if and only if $\hat{\theta}=\theta$.
The general form for weakly centered tests is
\begin{equation}
T_\Gamma=\frac{\gamma+\sum_{i=1}^{n-1}\Gamma_i^{(1)}\xi_i+\sum_{i,j=1}^{n-1}\Gamma_{ij}^{(2)}\xi_i\xi_j+\sum_{i,j,k=1}^{n-1}\Gamma_{ijk}^{(3)}\xi_i\xi_j\xi_k+\cdots}{\sqrt{\var\left( {\gamma+\sum_{i=1}^{n-1}\Gamma_i^{(1)}\xi_i+\sum_{i,j=1}^{n-1}\Gamma_{ij}^{(2)}\xi_i\xi_j+\sum_{i,j,k=1}^{n-1}\Gamma_{ijk}^{(3)}\xi_i\xi_j\xi_k+\cdots} \right)} }
\end{equation}
with the condition
\begin{equation}
0=\gamma+\sum_{i=1}^{n-1}\Gamma_i^{(1)}\mu_i+\sum_{i,j=1}^{n-1}\Gamma_{ij}^{(2)}\mu_{ij}+\sum_{i,j,k=1}^{n-1}\Gamma_{ijk}^{(3)}\mu_{ijk}+\ldots
\end{equation}
where the $\Gamma_{ijk\ldots}$ are real symmetric weights, possibly dependent on $\theta$. We can simplify these expressions using the same notation as above, obtaining the simpler form
\begin{equation}
T_\Gamma=\frac{\gamma+\sum_\mathbf{I}\Gamma_\mathbf{I}^{(n_\mathbf{I})}(\xi\ldots\xi)_\mathbf{I}}{\sqrt{\var\left( {\gamma+\sum_\mathbf{I}\Gamma_\mathbf{I}^{(n_\mathbf{I})}(\xi\ldots\xi)_\mathbf{I}} \right)} }\label{formweak}
\end{equation}
with the condition
\begin{equation}
0=\gamma+\sum_\mathbf{I}\Gamma_\mathbf{I}^{(n_\mathbf{I})}\mu_\mathbf{I}\label{condweak}
\end{equation}
Also for this class of tests, the sum can be limited to polynomials of fixed order or extended to power series. Note that the rescaled version of the $G$ test by Fu presented above belongs to this class.

The important difference between strongly and weakly centered tests is related to the robustness with respect to a biased estimation of $\theta$. Since the class of weakly centered tests is much larger than the class of strongly centered ones, it should be easier to find powerful tests in the former class than in the latter. However,  even if weakly centered tests could be more powerful, they would not be centered in scenarios where the value of $\theta$ could not be estimated precisely. On the other side, strongly centered tests are robust with respect to a bad estimation of $\theta$ and therefore they would be preferable in scenarios where 
an unbiased estimation of $\theta$ is troublesome. 

The scaling rule (\ref{scaling}) can be generalized to nonlinear tests in terms of functions $\w^{(n_\mathbf{I})}(f_1,f_2\ldots f_{n_\mathbf{I}})$ for strongly centered and $\Gamma^{(n_\mathbf{I})}(f_1,f_2\ldots f_{n_\mathbf{I}})$ for weakly centered tests:
\begin{equation}
\w_\mathbf{I}^{(n_\mathbf{I})}\simeq \frac{1}{n^{n_\mathbf{I}}}\w^{(n_\mathbf{I})}\left(\frac{i}{n},\frac{j}{n},\frac{k}{n}\ldots\right)\label{scalingnlin}
\end{equation}
\begin{equation}
\Gamma_\mathbf{I}^{(n_\mathbf{I})}\simeq \frac{1}{n^{n_\mathbf{I}}}\Gamma^{(n_\mathbf{I})}\left(\frac{i}{n},\frac{j}{n},\frac{k}{n}\ldots\right)\label{scalingnlin2}
\end{equation}
Fixing the precise scaling is more ambiguous than in the linear case because there are many different ways to preserve centeredness. For this reason the choice of scaling would be different for strongly and weakly centered tests and will not be discussed here.
  
All the possible nonlinear neutrality tests based on the frequency spectrum fall into one of the two classes presented in this section and have the form (\ref{formstrong}),(\ref{condstrong}) or (\ref{formweak}),(\ref{condweak}). Since both these classes contain an infinite number of possible choices of weights, the only reasonable criterion to study general nonlinear tests is to select the most powerful or interesting ones. Apart from the Hotelling choice of Fu \cite{fu1996new}, the most interesting choice is apparently the subclass of nonlinear optimal tests, which will be discussed in the next sections. 
 
\subsection{Strongly centered optimal tests}

As discussed for the linear case, optimal tests depend on the expected alternative scenario. In the nonlinear case, in principle it would be possible to find generic optimal tests, but there is no clear framework to obtain them. For  this reason we limit our study to the case of optimal tests for a specific alternative scenario. We denote by $\bar{\mu}_{ijk\ldots}=\mathcal{E}(\xi_i\xi_j\xi_k\ldots)$ the moments of the alternative spectrum for this scenario. 

Since we use the same normalization for linear and nonlinear tests, the optimality condition corresponds to the maximization of the expected value of the test under the alternative scenario
\begin{equation}
\mathcal{E}(T_\w)=\frac{\sum_{\mathbf{I}}\w_\mathbf{I}^{(n_\mathbf{I})}\bar{\mu}_{\mathbf{I}}}{\sqrt{\var\left( {\sum_{\mathbf{I}}\w_\mathbf{I}^{(n_\mathbf{I})}(\xi\ldots\xi)_{\mathbf{I}}} \right)}}\label{maxeqstrong}
\end{equation}
and can be justified as in the linear case. 



We denote by $\mathbf{\tilde{I}}$ the ordered sequence of the indices contained in $\mathbf{{I}}=ijk\ldots$ and by $\sigma\left(\mathbf{{I}}\right)$ the number of distinct permutations of the sequence $\mathbf{{I}}$, i.e. the total number of permutations divided by the number of permutations that leave $\mathbf{{I}}$ invariant. The main result for the optimal weights is presented in this theorem.
\begin{theo}\label{theosolstrong} The maxima of $\mathcal{E}(T_\w)$ correspond to the weights
\begin{equation}
\w_\mathbf{I}^{(n_\mathbf{{I}})}= 
\frac{1}{\sigma\left(\mathbf{{I}}\right)}\left( \sum_{\mathbf{\tilde{L}}}C^{-1}_{\mathbf{\tilde{I}}\mathbf{\tilde{L}}} \bar{\mu}_{\mathbf{\tilde{L}}} 
-
\sum_{k}\sum_{l} \sum_\mathbf{\tilde{L}} C^{-1}_{\mathbf{\tilde{I}}\mathbf{\tilde{L}}}\mu_\mathbf{\tilde{L}}^{(k)} \mathcal{M}_{kl} \sum_{\mathbf{\tilde{J}},\mathbf{\tilde{K}}}      
\mu_\mathbf{\tilde{J}}^{(l)}
C^{-1}_{\mathbf{\tilde{J}}\mathbf{\tilde{K}}} \bar{\mu}_{\mathbf{\tilde{K}}}
\right) \label{opts}
\end{equation} 
where the matrices $C^{-1}_{\mathbf{\tilde{I}}\mathbf{\tilde{J}}}$ and $\mathcal{M}_{kl}$ satisfy the identities
\begin{equation}
\sum_\mathbf{\tilde{K}}C^{-1}_{\mathbf{\tilde{I}}\mathbf{\tilde{K}}}\left({\mu}_{\mathbf{\tilde{K}}\mathbf{\tilde{J}}}-\mu_\mathbf{\tilde{K}}\mu_\mathbf{\tilde{J}}\right)=\delta_{\mathbf{\tilde{I}}\mathbf{\tilde{J}}}\label{matc}
\end{equation}
\begin{equation}
\sum_{r}\mathcal{M}_{kr}\sum_{\mathbf{\tilde{I}},\mathbf{\tilde{L}}}\mu_{\mathbf{\tilde{I}}}^{(r)}C^{-1}_{\mathbf{\tilde{I}}\mathbf{\tilde{L}}} \mu_{\mathbf{\tilde{L}}}^{(l)}=\delta_{kl}\label{mdefs}
\end{equation}
Moreover, the variance of the corresponding unnormalized test under the null model is equal to its expected value under the alternative model:
\beq
 \var\left( {\sum_{\mathbf{I}}\w_\mathbf{I}^{(n_\mathbf{I})}(\xi\ldots\xi)_{\mathbf{I}}} \right) =
\sum_{\mathbf{{I}}}\w_\mathbf{{I}}^{(n_\mathbf{I})}\bar{\mu}_{\mathbf{{I}}} 
\eeq
\end{theo}

Note that in general all the weights of the above optimal solution (\ref{opts}) are nonzero, therefore the maximum average value of the test for optimal tests built on polynomials of degree $d$ increases with the degree $d$. This suggests that optimal tests of higher degree should be more powerful than linear optimal tests. 

We provide explicit formulae for the above weights for the optimal quadratic test in the independent sites approximation. 
Given $E(\xi_i)=\mu_i$ and $\mathcal{E}({\xi}_i)=\bar{\mu}_i$, the relevant weights $\w_{\mathbf{I}}^{(n_\mathbf{I})}$ are  
\begin{align}
 \Omega_i^{(1)}&=  (\smo+2-\smb) \left(\frac{\bar{\mu}_i}{\mu_i}- \frac{\smb}{{\smo}}\right)- \frac{1}{2}\left(\frac{\bar{\mu}_i^2}{\mu_i^2}-\frac{ \smb^{2}}{\smo^2}\right) \label{scquad1}\\
 \Omega_{ii}^{(2)}&= -\left(\frac{\bar{\mu}_i}{\mu_i}- \frac{\smb}{{\smo}}\right) +\frac{1}{2}\left(\frac{\bar{\mu}_i^2}{\mu_i^2}-\frac{ \smb^{2}}{\smo^2}\right) \label{scquad2}\\
 \Omega_{ij}^{(2)}&= \frac{1}{2} \left[\left(\frac{\bar{\mu}_i\bar{\mu}_j}{\mu_i\mu_j}-\frac{ \smb^{2}}{\smo^2}\right) - \left(\frac{\bar{\mu}_i}{\mu_i}- \frac{\smb}{{\smo}}\right) - \left(\frac{\bar{\mu}_j}{\mu_j}- \frac{\smb}{{\smo}}\right)  \right]\label{scquad3}
\end{align}
where $\smo=\sum_{i=1}^{n-1}\mu_i$ and $\smb=\sum_{i=1}^{n-1}\bar{\mu}_i$. 
All these formulae are also valid for the folded spectrum if the appropriate $\mu_i$ and $\bar{\mu}_i$ are used. These results are discussed in Appendix \ref{moments}.

For optimal tests of higher degree, explicit expressions become cumbersome and the numerical implementation of the test (\ref{opts}) and the matrices (\ref{matc}), (\ref{mdefs}) is more convenient.

\subsection{Weakly centered optimal tests}
In this case the optimality condition corresponds to the maximization of the expression
\begin{equation}
\mathcal{E}(T_\Gamma)=\frac{\gamma+\sum_\mathbf{I}\Gamma_\mathbf{I}^{(n_\mathbf{I})}\bar{\mu}_{\mathbf{I}}}{\sqrt{\var\left( {\gamma+\sum_\mathbf{I}\Gamma_\mathbf{I}^{(n_\mathbf{I})}(\xi\ldots\xi)_\mathbf{I}} \right)} }\label{maxeqweak}
\end{equation}
with the same condition
\begin{equation}
0=\gamma+\sum_\mathbf{I}\Gamma_\mathbf{I}^{(n_\mathbf{I})}\mu_\mathbf{I}\label{condweak2}
\end{equation}
The simplest case corresponds to a first order polynomial
\begin{equation}
T_\Gamma=\frac{\gamma+\sum_{i=1}^{n-1}\Gamma_i^{(1)}\xi_i}{\sqrt{ \gamma^2+2\gamma\sum_{j=1}^{n-1}\Gamma_j^{(1)}\mu_j+\sum_{j=1}^{n-1}\sum_{k=1}^{n-1}\Gamma_j^{(1)}\Gamma_k^{(1)} \mu_{jk}}}\label{wc1order}
\end{equation}
whose maximum corresponds to the optimal weights
\begin{equation}
\Gamma_i^{(1)}=\sum_{j=1}^{n-1}c^{-1}_{ij}\left(\bar{\mu}_j-\mu_j\right)\quad ,\quad  \gamma=-\sum_{j=1}^{n-1}\sum_{k=1}^{n-1}\mu_jc^{-1}_{jk}\left(\bar{\mu}_k-\mu_k\right)
\end{equation}
where $c^{-1}_{ij}$ is the inverse matrix of the covariance matrix $c_{ij}=\mu_{ij}-\mu_{i}\mu_{j}$. 
Since $\gamma\neq 0$ for this optimal test, the value of this test for the specific scenario for which it is built is larger than than the value of the corresponding linear optimal test. 
In fact the maximum of the test is
\begin{equation}
\mathcal{E}(T_\Gamma)=\sqrt{\sum_{j=1}^{n-1}\sum_{k=1}^{n-1}\left(\bar{\mu}_j-\mu_j\right) c^{-1}_{jk}\left(\bar{\mu}_k-\mu_k\right)}
\end{equation}
that should be compared to the maximum of the optimal test for the linear case, which can be rewritten as 
\begin{equation}
\mathcal{E}(T_\w)_{linear}=\sqrt{\sum_{j=1}^{n-1}\sum_{k=1}^{n-1}\left(\bar{\mu}_j-\mu_j\right) c^{-1}_{jk}\left(\bar{\mu}_k-\mu_k\right)- \frac{\left(\sum_{j=1}^{n-1}\sum_{k=1}^{n-1}\mu_j c^{-1}_{jk}\left(\bar{\mu}_k-\mu_k\right)\right)^2}
{ \sum_{j=1}^{n-1}\sum_{k=1}^{n-1}\mu_j c^{-1}_{jk}\mu_k}}
\end{equation}
The comparison shows clearly that nonlinear optimal tests are always more powerful than linear optimal tests for the same scenario.

The form of the results for the general case is similar to this simple case.
\begin{theo}\label{theosolweak} The maxima of $\mathcal{E}(T_\Gamma)$ correspond to the weights
\begin{equation}
\Gamma_\mathbf{I}^{(n_\mathbf{{I}})}=\frac{1}{\sigma\left(\mathbf{{I}}\right)} \sum_{\mathbf{\tilde{J}}}C^{-1}_{\mathbf{\tilde{I}}\mathbf{\tilde{J}}}\left(\bar{\mu}_\mathbf{\tilde{J}}-\mu_\mathbf{\tilde{J}}\right)\quad ,\quad  \gamma=-\sum_\mathbf{\tilde{J}}\mu_\mathbf{\tilde{J}}\sum_\mathbf{\tilde{K}}C^{-1}_{\mathbf{\tilde{J}}\mathbf{\tilde{K}}}\left(\bar{\mu}_\mathbf{\tilde{K}}-\mu_\mathbf{\tilde{K}}\right)\label{optw}
 \end{equation}
where $C^{-1}_{\mathbf{\tilde{I}}\mathbf{\tilde{J}}}$ satisfied the identity
\begin{equation}
\sum_\mathbf{\tilde{K}}C^{-1}_{\mathbf{\tilde{I}}\mathbf{\tilde{K}}}\left({\mu}_{\mathbf{\tilde{K}}\mathbf{\tilde{J}}}-\mu_\mathbf{\tilde{K}}\mu_\mathbf{\tilde{J}}\right)=\delta_{\mathbf{\tilde{I}}\mathbf{\tilde{J}}}
 \end{equation}
Moreover, the variance of the corresponding unnormalized test under the null model is equal to its expected value under the alternative model:
\beq
  \var\left( {\gamma+\sum_\mathbf{I}\Gamma_\mathbf{I}^{(n_\mathbf{I})}(\xi\ldots\xi)_\mathbf{I}} \right) =
\sum_\mathbf{{I}}\Gamma_\mathbf{{I}}^{(n_\mathbf{I})}\bar{\mu}_{\mathbf{{I}}}+\gamma
\eeq
\end{theo}

Also in this case, the power of optimal tests based on polynomials of higher degree increases with the degree of the polynomial.

It is possible to give explicit expressions of the above matrix and moments for the optimal quadratic test. The formulae for the weights $\Gamma_{\mathbf{I}}^{(n_\mathbf{I})}$ for the unfolded spectrum are 
\begin{align}
 \Gamma_i^{(1)} &=  (\smo+2-\smb) \left(\frac{\bar{\mu}_i}{\mu_i}-1\right)- \frac{1}{2}\left(\frac{\bar{\mu}^2_i}{\mu_i^2}-1\right) \label{wcquad1}\\
 \Gamma_{ii}^{(2)} & =  \frac{1}{2} \left(\frac{\bar{\mu}_i}{\mu_i}-1\right)^2   \label{wcquad2}\\
 \Gamma_{ij}^{(2)} & = \frac{1}{2} \left(\frac{\bar{\mu}_i}{\mu_i}-1\right)\left(\frac{\bar{\mu}_j}{\mu_j}-1\right) \label{wcquad3}\\
\gamma & = \frac{1}{2} (\smo-\smb)(\smo+2-\smb) \label{wcquad4}
\end{align}
These results are valid in the independent sites approximation. They are also valid for the folded spectrum if the appropriate $\mu_i$ and $\bar{\mu}_i$ are used. An expression for the denominator of the test in the independent sites approximation can be found in Appendix \ref{moments}.


\subsection{Simulations of the power of optimal tests}

Since a theoretical evaluation of the power of optimal tests of different degree is not possible, we evaluate numerically the power of some of these tests in different scenarios. We consider the best possible case, that is, we assume that the precise value of $\theta$ is known. Moreover we assume unlinked sites and $\theta\ll 1$. In this approximation, as shown in Appendix \ref{moments}, the moments $E(\xi_i\xi_j\xi_k\ldots)$ depend only on the first moments $\mu_i=\theta L \xz_i$ and similarly $\mathcal{E}(\xi_i\xi_j\xi_k\ldots)$ depend only on $\bar{\mu}_i=\theta L \xb_i$, therefore optimal tests depend only on the alternative and null average spectra.


Note that for numerical simulations of optimal tests of higher degree, the numerical implementation can be made easier if all the
occurrences
of inverse covariance matrices $C^{-1}_{\mathbf{\tilde{I}}\mathbf{\tilde{J}}}$ in the the above formulae are replaced with the corresponding
second moments $\mu^{-1}_{\mathbf{\tilde{I}}\mathbf{\tilde{J}}}$, both in the expressions (\ref{opts}), (\ref{optw}) and in the definition
(\ref{mdefs}). The test is the same because of the centeredness condition, as it can be verified explicitly.



We compare four optimal tests. The first two are the linear and quadratic strongly centered optimal tests, which are denoted by $T_{O(1)}^{sc}$ and $T_{O(2)}^{sc}$ respectively. The third test is the weakly centered optimal test $T_{O(1)}^{wc}$ based on a first order polynomial and presented in (\ref{wc1order}). The last optimal test $T_{O(2)}^{wc}$ is also weakly centered and based on on a quadratic polynomial. The explicit formulae for the computation of the weights of $T_{O(2)}^{sc}$ and $T_{O(2)}^{wc}$ were given in equations (\ref{scquad1})-(\ref{scquad3}) and (\ref{wcquad1})-(\ref{wcquad4}). 

We simulated two demographic processes: (A) subdivision, where two populations having identical size exchange individuals given a symmetric migration rate $M$, then individuals are sampled from one population only; (B) expansion, where the population size changes by a factor $N_0/N=10$ at a time $T$ before present (in units of $4N$ generations). For each value of the parameters $M$ and $T$, $10^6$ simulations were performed with \textit{mlcoalsim} v1.98b \cite{Ramos-Onsins2007a} for a region of 1000 bases with variability $\theta=0.05$ and recombination $\rho=\infty$ and a sample size of $n=20$ (haploid) individuals. Confidence intervals at 95\% level were estimated from $10^6$ simulations of the standard neutral coalescent with the same parameters.

\begin{center}
\begin{figure}
 \includegraphics[scale=0.325]{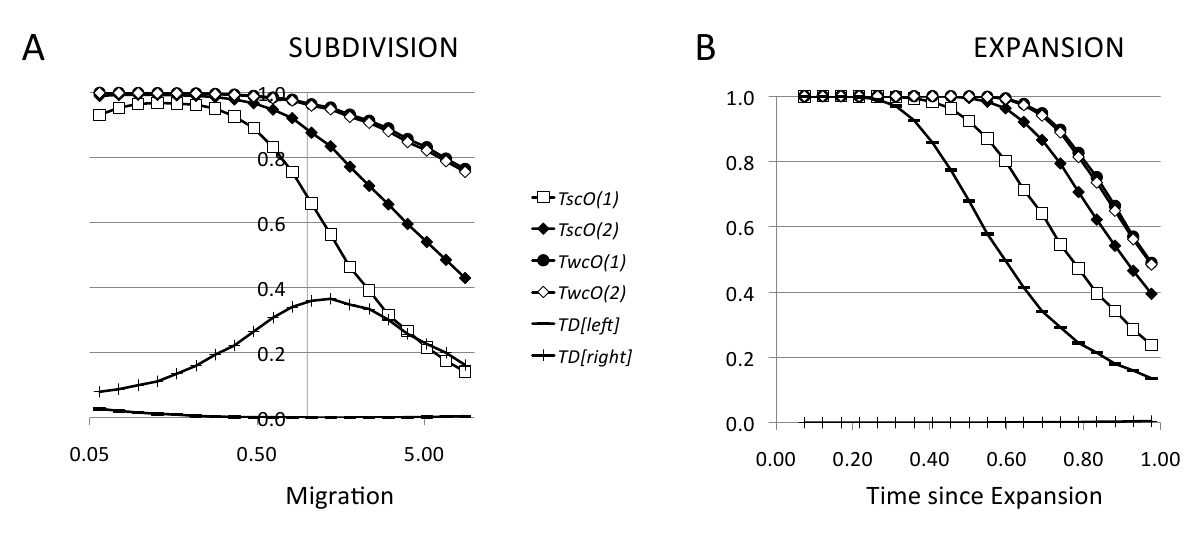}
\caption{Statistical power of nonlinear optimal tests from coalescent simulations for the 5\% tail, compared with Tajima's $D$ test
(for the left and the right tail). The parameters for the simulations are: $n=20$, $\theta=0.05$, $L=1000$bp, $\rho=\infty$; two
populations considered but only one sampled (for panel A); 
expansion factor $N_0/N=10$ (for panel B).}
\label{fig_test}
\end{figure}
\end{center}

In Figure \ref{fig_test} we compare the power of the tests in the best possible situation, namely when $\theta$ is known with good precision. In this condition all optimal tests should give the best results. In fact, the power of weakly centered tests ($T_{O(1)}^{wc}$ and $T_{O(2)}^{wc}$) is impressive, being around $100\%$ for a large part of the parameter space and decreasing for large migration rates (Figure \ref{fig_test}A) and long times (Figure \ref{fig_test}B) as every other test, because the frequency spectrum for these cases becomes very similar to the standard spectrum. So, weakly centered tests show a very good theoretical performance, counterbalanced by their lack of robustness. The power of $T_{O(1)}^{wc}$ and $T_{O(2)}^{wc}$ are almost identical, therefore the contribution of the quadratic part to $T_{O(2)}^{wc}$ is probably not relevant. 

On the other hand, strongly centered optimal tests are quite more powerful than Tajima's $D$ but less powerful than weakly centered
tests, as expected. However, there is a sensible difference in power between $T_{O(1)}^{sc}$ and $T_{O(2)}^{sc}$: in the range of
parameters where the power of weakly centered tests is around $100\%$, both strongly centered tests show a good performance not so far
from the weakly centered ones, while in the less favourable range the quadratic test $T_{O(2)}^{sc}$, while performing worse than the
weakly centered tests, has a power that is  $20\%$ higher than the linear test $T_{O(1)}^{sc}$. Taking into account the robustness of
the tests, these simulations show that optimal tests like $T_{O(2)}^{sc}$ could be an interesting alternative to the usual linear
tests.
 


\section{Conclusions}

In this paper we have presented a systematic analysis of neutrality tests based on the site frequency spectrum. This study is intended to extend and complete the recent works in \cite{achaz2009frequency} and \cite{ferretti2010optimal} by extending the study of the linear neutrality tests recently presented by Achaz, their properties and the optimal tests that can be obtained in this framework; in a further generalization, we also consider the most general class of tests that can be written as a power series of the different components $\xi_i$ of the spectrum. The aim of the paper is to give mathematical guidelines to build new and more effective tests. The proposed guidelines are the scaling relation (\ref{scaling}) and the optimality condition based on the maximization of $\mathcal{E}(T_\Omega)$. Both these guidelines are thoroughly explained and discussed.

While nonlinear optimal tests have been shown to be more powerful than linear ones (and weakly centered tests more powerful than strongly
centered ones), power is not the only important issue: also robustness must be taken into consideration. In fact there are three important
remarks on the relative robustness of these tests. The first one is that, as already discussed, centeredness of weakly centered tests is not
robust with respect to a biased estimate of $\theta$, therefore these tests should be preferred to strongly centered tests only in situations where the
value of $\theta$ is well known or a good estimate is available.

The second remark is that neither the weights nor the results of linear optimal tests do depend on the value of $\theta$ and on the
number of segregating sites $S$, while the weights of nonlinear optimal tests have an explicit dependence on $\theta$ and their results
depend not only on the spectrum but also on $S$, therefore the interpretation of the results of these tests is more complicated.
However, this is not necessarily true for homogeneous tests of any degree, like the quadratic $G_\xi$ test by Fu. An interesting
development of this work could be a study of  homogeneous tests of a given degree $k$ satisfying the optimality condition, which can be
easily obtained from equations (\ref{opts}) by restricting all ordered sequences of indices
$\mathbf{\tilde{I}},\mathbf{\tilde{J}},\mathbf{\tilde{K}},\mathbf{\tilde{L}}$ to contain precisely $k$ indices (along with some
``traceless'' condition, in case). These homogeneous optimal tests (or at least some subclass of them) should depend weakly on $S$. 

The third remark is that linear optimal tests have two interesting properties that are not shared by other tests: they depend only on the deviations from the null spectrum and they have an easy interpretation in terms of these deviations, that is, they are positive if the observed deviations are similar to the expected ones and negative if the observed deviations are opposite to the expected ones. These features give an important advantage to linear optimal tests. 

Tests based on the frequency spectrum of polymorphic sites are fast, being based on simple matrix multiplications, and can be therefore applied to genome-wide data. Moreover, they can be used as summary statistics for Approximate Bayesian Computation or other statistical approaches to the analysis of sequence data. While linear tests are often used in this way, the nonlinear tests presented in this paper contain more information (related to the covariances and higher moments of the frequency spectrum) that could increase the power of these analyses.



\subsection*{Acknowledgments:} 
We thank M. Perez-Enciso and J. Rozas for useful comments.      
 G.M. acknowledges support by Fondazione A. Della Riccia and Japan Society for Promotion
of Science. Work funded by grant CGL2009-0934 (MICINN,
Spain) to S.R.O. 
\begin{appendices}
\section{Moments of the frequency spectrum in the independent sites approximation}\label{moments}

We consider the limit $\theta\ll 1$, $L\rightarrow\infty$ with constant $\theta L$. The spectrum $\xi_i$ can be written as a sum of spectra for all sites 
\beq
\xi_i=\sum_{s=1}^L\xi_i(s)
\eeq
where each variable $\xi_i(s)$ has a Bernoulli distribution $\xi_i(s)\in\{0,1\}$ with probabilities $p(1)=\theta\xz_i$ and $p(0)=1-\theta\xz_i$ where $\xz_i=1/i$ under the standard neutral model. 
The expectation value of $\xi_i$ is therefore
\beq
E(\xi_i)=\mu_i=\sum_{s=1}^LE(\xi_i(s))=\theta L\xz_i=\frac{\theta L}{i}
\eeq
and similarly $\mathcal{E}(\xi_i)=\bar{\mu}_i=\theta L\xb_i$ for a general model with average spectrum $\xb_i$. 

In the independent sites approximation, which is equivalent to the infinite recombination limit, the variables  $\xi_i(s)$ and $\xi_i(s')$ are 
i.i.d. random variables for $s\neq s'$, and more generally the random variables  $\xi_i(s)$ and $\xi_j(s')$ are independent for $s\neq s'$. The moments for a single site $s$ can be calculated as
\begin{align}
E(\xi_i(s)\xi_j(s)\xi_k(s)\ldots)=&\sum_{a,b,c\ldots\in\{0,1\}}abc\ldots P(\xi_i(s)=a,\xi_j(s)=b,\xi_k(s)=c,\ldots)=\nonumber\\
=P(\xi_i(s)=1,\xi_j(s)=&1,\xi_k(s)=1,\ldots)=P(\xi_i(s)=1)\delta_{ij}\delta_{jk}\ldots 
\end{align}
because the sum $\sum_{i=1}^{n-1}
\xi_i(s)\in\{0,1\}$, that is, different allele frequencies for the same site are mutually exclusive. Therefore all these moments are always linear in $\theta$ and are nonzero only when all indices are equal:
\beq
E(\xi_i(s)\xi_j(s)\xi_k(s)\ldots)=\theta\xz_i\delta_{ij}\delta_{jk}\ldots=\frac{\theta}{i}\delta_{ij}\delta_{jk}\ldots\quad .
\eeq

Then the second moment can be evaluated as
\begin{align}
E(\xi_i\xi_j)=&\sum_{s,s'}E(\xi_i(s)\xi_j(s'))=\sum_sE(\xi_i(s)\xi_j(s))+\sum_{s\neq s'}E(\xi_i(s))E(\xi_j(s'))= \nonumber\\
=&L\delta_{ij}\frac{\theta }{i} + L(L-1)\frac{\theta^2}{ij}=\delta_{ij}\mu_i+\mu_i\mu_j-\frac{\mu_i\mu_j}{L}
\end{align}
and neglecting subleading orders in $\theta$ or $L^{-1}$ like the last term above, we can calculate the third and forth moments that are needed for the calculation of $C^{-1}$. 

The final result for the moments of the spectrum $\xi_i$ is
\begin{align}
E(\xi_i\xi_j)=\mu_{ij}=& \delta_{ij}\mu_i+\mu_i\mu_j \\
E(\xi_i\xi_j\xi_k)=\mu_{ijk} =& \delta_{ij}\delta_{jk}\mu_i+(\delta_{ik}\mu_i\mu_j+\delta_{jk}\mu_i\mu_j+\delta_{ij}\mu_i\mu_k)+\mu_i\mu_j\mu_k \\
E(\xi_i\xi_j\xi_k\xi_l)= \mu_{ijkl}=& \delta_{ij}\delta_{jk}\delta_{kl}\mu_i+(\delta_{ik}\delta_{jl}\mu_i\mu_j+\delta_{il}\delta_{jk}
\mu_i\mu_j+\delta_{ij}\delta_{kl}\mu_i\mu_k)+ \nonumber\\
+& (\delta_{ij}\delta_{jk}\mu_i\mu_l+\delta_{ij}\delta_{jl}\mu_i\mu_k+\delta_{ik}\delta_{kl}\mu_i\mu_j+\delta_{jk}\delta_{kl}
\mu_i\mu_j)+ \nonumber\\
+& (\delta_{il}\mu_i\mu_j\mu_k+\delta_{jl}\mu_i\mu_j\mu_k+\delta_{ik}\mu_i\mu_j\mu_l+\delta_{jk}\mu_i\mu_j\mu_l+\nonumber \\
+&\delta_{ij}\mu_i\mu_k\mu_l+\delta_{kl}\mu_i\mu_j\mu_k)+\mu_i\mu_j\mu_k\mu_l
\end{align}

All the results above can be applied to a general model simply by substituting $\mu_i$ with $\bar{\mu}_i$. Moreover they can be applied to the folded spectrum by taking $\mu_i=\theta L {n}/{i(n-i)(1+\delta_{n,2i})}$ for the standard neutral model or $\bar{\mu}_i={\theta L (\xb_i+\xb_{n-i})}/{(1+\delta_{n,2i})}$ for general models.

We define some quantities in order to simplify the expressions for the weights:
\begin{align}
\smo &=\sum_{i=1}^{n-1}\mu_i \quad, &\quad
\smb &=\sum_{i=1}^{n-1}\bar{\mu}_i \quad, &\quad
\smq &=\sum_{i=1}^{n-1}\frac{\bar{\mu}_i^2}{\mu_i}
\end{align}
If the spectrum is folded, all the sums in the above expressions run from $1$ to $\fln$.

The covariance matrix $C_{\mathbf{\tilde{I}},\mathbf{\tilde{J}}}$ is
\begin{align}
C_{i,j} &=\mu_{ij}-\mu_i\mu_j=\delta_{ij}\mu_i \\
C_{ij,k} &= \mu_{ijk}-\mu_{ij}\mu_k=\delta_{ij}\delta_{jk}\mu_i+(\delta_{ik}\mu_i\mu_j+\delta_{jk}\mu_i\mu_j) \\
C_{ij,kl} &= \mu_{ijkl}-\mu_{ij}\mu_{kl}=\delta_{ij}\delta_{jk}\delta_{kl}\mu_i+(\delta_{ik}\delta_{jl}\mu_i\mu_j+\delta_{il}\delta_{jk}
\mu_i\mu_j)+ \nonumber\\
& +(\delta_{ij}\delta_{jk}\mu_i\mu_l+\delta_{ij}\delta_{jl}\mu_i\mu_k+\delta_{ik}\delta_{kl}\mu_i\mu_j+\delta_{jk}\delta_{kl}
\mu_i\mu_j)+ \nonumber\\
& +(\delta_{il}\mu_i\mu_j\mu_k+\delta_{jl}\mu_i\mu_j\mu_k+\delta_{ik}\mu_i\mu_j\mu_l+\delta_{jk}\mu_i\mu_j\mu_l)
\end{align}
with the elements $C_{ij,k}$ and $C_{ij,kl}$ that should be considered only for $i\leq j$, $k\leq l$. It can be verified that the inverse matrix $C^{-1}_{\mathbf{\tilde{I}},\mathbf{\tilde{J}}}$ has the form 
\begin{align}
 C_{i,j}^{-1} =& 1+\delta_{ij}\frac{2\mu_i\left(\smo+3\right)+1}{2\mu_i^2}\label{invmat1}\\
C_{ij,k}^{-1} =& \delta_{ij}\delta_{jk}\frac{2\mu_i-1}{2\mu_i^2}-(\delta_{ik}+\delta_{jk})\frac{1}{\mu_k} \label{invmat2} \\
C_{ij,kl}^{-1} =& (\delta_{ik}\delta_{jl}+ \delta_{il}\delta_{jk})\frac{1}{\mu_i\mu_j}-\frac{3}{2}\delta_{ij}\delta_{ik}\delta_{jl}\frac{1}{\mu_i^2}\label{invmat3}
\end{align}

The formulae for the weakly centered quadratic test (\ref{wcquad1}-\ref{wcquad4}) can be obtained from these formulae and the definition (\ref{optw}). The corresponding variance in the denominator  (\ref{formweak}) is 
\begin{eqnarray}
{{\var\left( {\sum_\mathbf{I}\Gamma_\mathbf{I}^{(n_\mathbf{I})}(\xi\ldots\xi)_\mathbf{I}} \right)} }&=&
2(\smb-\smq/2)^2 +\smo(\smo/2+1-2\smb+\smq) 
\nonumber\\ && +\smq-2\smb
\end{eqnarray}

The matrix $\mathcal{M}$ is the inverse of the matrix $\mathcal{M}^{-1}_{rl}=\sum_{\mathbf{\tilde{I}},\mathbf{\tilde{L}}}\mu_{\mathbf{\tilde{I}}}^{(r)}C^{-1}_{\mathbf{\tilde{I}}\mathbf{\tilde{L}}} \mu_{\mathbf{\tilde{L}}}^{(l)}$, which can be easily calculated from the above equations as
\begin{align}
\mathcal{M}^{-1}_{11} &= 2 \smo^2+\smo\\
\mathcal{M}^{-1}_{12} &= - \smo^2\\ 
\mathcal{M}^{-1}_{22} &= \smo^2/2
\end{align}
and the matrix $\mathcal{M}$ is
\begin{equation}
 \mathcal{M}={
\begin{pmatrix} 1/\smo &2/\smo\\ 
 2/\smo &  4/\smo+2/\smo^{2}\end{pmatrix} }\label{matm}
 \end{equation}
The formulae (\ref{scquad1}-\ref{scquad3}) can be obtained from the formula (\ref{matm}) and from the following results:
\begin{align}
 \sum_\mathbf{\tilde{I}} C^{-1}_{i,\mathbf{\tilde{I}}} \bar{\mu}_\mathbf{\tilde{I}} &= \smb + (\smo+2-\smb) \frac{\bar{\mu}_i}{\mu_i} -\frac{1}{2} \left(\frac{\bar{\mu}_i}{\mu_i}\right)^2 \\
 \sum_\mathbf{\tilde{I}}C^{-1}_{ii,\mathbf{\tilde{I}}} \bar{\mu}_\mathbf{\tilde{I}} &= -\frac{\bar{\mu}_i}{\mu_i} + \frac{1}{2} \left(\frac{\bar{\mu}_i}{\mu_i}\right)^2 \\
 \sum_\mathbf{\tilde{I}}C^{-1}_{ij,\mathbf{\tilde{I}}} \bar{\mu}_\mathbf{\tilde{I}} &= - \frac{\bar{\mu}_i}{\mu_i} - \frac{\bar{\mu}_j}{\mu_j} + \frac{\bar{\mu}_i\bar{\mu}_j}{\mu_i\mu_j}
\end{align}
\begin{align}
\sum_{\mathbf{\tilde{I}},\mathbf{\tilde{J}}}\mu_{\mathbf{\tilde{I}}}^{(1)} C^{-1}_{\mathbf{\tilde{I}}\mathbf{\tilde{J}}} \bar{\mu}_\mathbf{\tilde{J}} &= \smb(2\smo-\smb+1) \quad &
\sum_{\mathbf{\tilde{I}},\mathbf{\tilde{J}}}\mu_{\mathbf{\tilde{I}}}^{(2)} C^{-1}_{\mathbf{\tilde{I}}\mathbf{\tilde{J}}} \bar{\mu}_{\mathbf{\tilde{J}}} &=\smb^2/2 - \smo \smb
\end{align}
\begin{align}
 \sum_\mathbf{\tilde{I}}C^{-1}_{i,\mathbf{\tilde{I}}} \mu^{(1)}_\mathbf{\tilde{I}} &= 2\smo + 2 \quad &
 \sum_\mathbf{\tilde{I}}C^{-1}_{i,\mathbf{\tilde{I}}} \mu^{(2)}_\mathbf{\tilde{I}} &= -\smo -1/{2}\\
 \sum_\mathbf{\tilde{I}}C^{-1}_{ii,\mathbf{\tilde{I}}} \mu^{(1)}_{\mathbf{\tilde{I}}} &=-1 \quad &
 \sum_\mathbf{\tilde{I}}C^{-1}_{ii,\mathbf{\tilde{I}}} \mu^{(2)}_{\mathbf{\tilde{I}}} &= 1/{2} \\
 \sum_\mathbf{\tilde{I}}C^{-1}_{ij,\mathbf{\tilde{I}}} \mu^{(1)}_{\mathbf{\tilde{I}}} &=-2 \quad &
 \sum_\mathbf{\tilde{I}}C^{-1}_{ij,\mathbf{\tilde{I}}} \mu^{(2)}_{\mathbf{\tilde{I}}} &= 1
\end{align}

\section{Proofs}
\label{proofs}

\newproof{prove1}{Proof of theorem \ref{theo1}}
\begin{prove1}
Choose a vector $\Delta_i$ in $\mathbb{R}^{n-1}$ that is orthogonal both to $i\w_i$ and to $i$, that is, such that $\sum_ii\Delta_i\w_i=0$ and $\sum_ii\w_i=0$. Since $\alpha/{ia_n}+(1-\alpha)\Delta_i$ is a set of continuous functions of $\alpha$, its minimum is also continuous in $\alpha$. Moreover, the minimum is clearly positive if $\alpha=1$, while it is negative by construction if $\alpha=0$. The theorem follows from the intermediate value theorem. $\blacksquare$
\end{prove1}

\newproof{prove2}{Proof of theorem \ref{theo2}}
\begin{prove2}
The proof is similar to the previous one. Choose a function $\Delta(f)$ in $C^\infty(0,1)$ to satisfy both $\int_{1/N}^1df\, f \wi(f)\Delta(f)=0$ and $\int_{1/N}^1df\, f\Delta(f)=0$. (Since these conditions correspond just to two independent functionals of $\Delta(f)$ and $C^\infty(0,1)$ is an infinite-dimensional linear space, the existence of such a function is guaranteed.) Since $\alpha/f\log(N)+(1-\alpha)\Delta(f)$ is a continuous functions of $\alpha$ and its infimum is not $\pm\infty$, its infimum is also continuous in $\alpha$. Moreover, the infimum is clearly positive if $\alpha=1$, while it is negative by construction if $\alpha=0$. The theorem follows from the intermediate value theorem. $\blacksquare$
\end{prove2}

\newproof{prove3}{Proof of theorem \ref{theo3}}
\begin{prove3}
The vectors $\w_i$ are a basis of the subspace $\mathbb{R}^{n-2}\subset\mathbb{R}^{n-1}$ defined by the condition $\sum_i\w_i=0$, that is, the space of vectors orthogonal to the vector whose components are $v_i=1$. Therefore the only vectors $i\Delta_i$ that are orthogonal to all the vectors in this basis are precisely of the form $i\Delta_i\propto v_i$, that is, $\Delta_i=const/i$. $\blacksquare$
\end{prove3}

The theorems on the form of optimal tests can be easily proved from a general lemma.

\newtheorem{lemma}{Lemma} 
\begin{lemma}\label{gentheo}
Consider a function $f:\mathbb{R}^M\smallsetminus \{0\}\rightarrow\mathbb{R}$ of the form
\beq\label{genfunc}
f(\vec{v})=\frac{\vec{v}\cdot\vec{w}}{\sqrt{\vec{v}\cdot Q \vec{v}}}
\eeq
where $\vec{w}\in\mathbb{R}^M$ and $Q$ is a $M\times M$ symmetric positive matrix, and a $K\times M$ matrix $R$ with $K<M$ and maximum rank. The extrema of the function $f$ restricted to the subspace $R\vec{v}=0$ are given by 
\beq\label{gensol}
\vec{v}_\alpha=\alpha\left(Q^{-1}\vec{w}-Q^{-1}R^t\left(RQ^{-1}R^t\right)^{-1}RQ^{-1}\vec{w}\right)
\eeq
The extrema with $\alpha>0$ are maxima and the extrema with $\alpha<0$ are minima of the function $f$. These extrema satisfy the identity
\beq
\vec{v}_\alpha\cdot Q \vec{v}_\alpha=\alpha\vec{v}_\alpha\cdot\vec{w}\label{idvaropt}
\eeq
\end{lemma}
\begin{prove}
The existence of maxima and minima can be proved by the Weierstrass extreme value theorem. In fact $f$ is continuous and invariant under a homothety with center in the origin of $\mathbb{R}^M$ and positive scale factor, therefore the codomain of the function on the linear subspace defined by $R\vec{v}=0$ is the same as the codomain of its restriction to the submanifold of unit vectors $|\vec{v}|=1$, which is a compact space. The restriction of $f$ is also continuous and the conclusion follows. To determine the extrema, the method of Lagrange multipliers states that it is sufficient to extremize the function 
\beq\label{genfunc2}
F(\vec{v},\vec{\lambda})=\frac{\vec{v}\cdot\vec{w}}{\sqrt{\vec{v}\cdot Q \vec{v}}}+\vec{\lambda}\cdot R\vec{v}
\eeq
and since there are no boundaries, this is equivalent to the solution of the equations
\begin{align}
0=\vec{\nabla}_v f=&\frac{\vec{w}}{\sqrt{\vec{v}\cdot Q \vec{v}}}-\frac{\vec{v}\cdot\vec{w}}{\left(\vec{v}\cdot Q \vec{v}\right)^{3/2}} Q \vec{v} +R^t \vec{\lambda} \label{eqlagr}\\
0=\vec{\nabla}_\lambda f=&R\vec{v} \label{constraints}
\end{align}
The solution satisfies
\beq
\frac{\vec{v}\cdot\vec{w}}{{\vec{v}\cdot Q \vec{v}}} \vec{v}=Q^{-1}{\vec{w}} + Q^{-1}R^t \vec{\lambda}\label{eqlagr2}
\eeq
and multiplying it by $R$ and using (\ref{constraints}) we obtain
\beq
\vec{\lambda}=-\left(RQ^{-1}R^t\right)^{-1}RQ^{-1}\vec{w}+\vec{l}\quad, \quad R^t\vec{l}=0
\eeq
that can be inserted again in equation (\ref{eqlagr2}) to eliminate $\vec{\lambda}$. The resulting equation  in $\vec{v}$ admits only solutions of the form (\ref{gensol}) and by substituting (\ref{gensol}) into it, it can be checked that all values of $\alpha\neq 0$ correspond to solutions of (\ref{eqlagr}),(\ref{constraints}). The invariance of $f$ under a central homothety with positive scale factor implies that the value of the function does not depend on $|\alpha|$. The function is positive for $\alpha>0$ and negative for $\alpha<0$, therefore solutions with $\alpha>0$ correspond to maxima and solutions with $\alpha<0$ correspond to minima. The identity (\ref{idvaropt}) can be proved by substituting the solution (\ref{gensol}).
$\blacksquare$
\end{prove}

\newproof{provesolstrong}{Proof of theorem 
\ref{theosolstrong}}
\begin{provesolstrong}
The expected values of the tests (\ref{formstrong}) have the same functional form as the function $f$ of Lemma \ref{gentheo}. The correspondence is the following:
\begin{align}
 \vec{v}\ \rightarrow\ & v_\mathbf{\tilde{I}}=\sigma(\mathbf{\tilde{I}})\w^{(n_\mathbf{\tilde{I}})}_\mathbf{\tilde{I}} \\
\vec{w}\ \rightarrow\ &w_\mathbf{\tilde{I}}=\bar{\mu}_\mathbf{\tilde{I}}\\
Q\ \rightarrow\ & Q_{\mathbf{\tilde{I}}\mathbf{\tilde{J}}}= \mu_{\mathbf{\tilde{I}}\mathbf{\tilde{J}}}-\mu_{\mathbf{\tilde{I}}}\mu_{\mathbf{\tilde{J}}}\\
R\ \rightarrow\ & R_{k,\mathbf{\tilde{I}}}=\mu^{(k)}_\mathbf{\tilde{I}}
\end{align}
and the positivity of the matrix $Q$ is guaranteed by the positivity of the variance for all possible choices of the weights. Application of Lemma \ref{gentheo} with a$\alpha=1$ gives the result (\ref{opts}). $\blacksquare$
\end{provesolstrong}

\newproof{provesolweak}{Proof of theorem \ref{theosolweak}}
\begin{provesolweak}
We can immediately solve equation (\ref{condweak2}) for $\gamma$ and substitute it in equation (\ref{maxeqweak}). Then $\gamma$ is a function of the other weights and the maximization is unconstrained. It can be seen that also in this case, the expected values of the tests have the same functional form as the function $f$ of Lemma \ref{gentheo}. The correspondence is the following:
\begin{align}
 \vec{v}\ \rightarrow\ & v_\mathbf{\tilde{I}}=\sigma(\mathbf{\tilde{I}})\Gamma^{(n_\mathbf{\tilde{I}})}_\mathbf{\tilde{I}} \\
\vec{w}\ \rightarrow\ & w_\mathbf{\tilde{I}}=\bar{\mu}_\mathbf{\tilde{I}}- {\mu}_\mathbf{\tilde{I}} \\
Q\ \rightarrow\ & Q_{\mathbf{\tilde{I}}\mathbf{\tilde{J}}}={\mu}_{\mathbf{\tilde{I}}\mathbf{\tilde{J}}} -{\mu}_\mathbf{\tilde{I}}{\mu}_\mathbf{\tilde{J}} \\
R\ \rightarrow\ & \mathrm{empty}\ 0\times M\ \mathrm{matrix}
\end{align}
and the positivity of the matrix $Q$ is implied by by the positivity of the variance. Then the result (\ref{optw}) follows from Lemma \ref{gentheo} with $\alpha=1$. $\blacksquare$
\end{provesolweak}

\end{appendices}



\section*{References}

\bibliographystyle{elsarticle-num}
\bibliography{popgen}







\end{document}